\documentclass[a4paper,10pt]{elsarticle}

\usepackage{amsmath,amssymb}
\usepackage{graphicx,pstricks}
\usepackage {geometry}
\usepackage{mathenv}
\usepackage[colorlinks=true]{hyperref} 

\hypersetup{urlcolor=blue,linkcolor=blue,citecolor=blue,colorlinks=true} 

\usepackage[toc,page]{appendix}

\newcommand{\eref}[1]{Eq. (\ref{#1})}
\newcommand{\efig}[1]{Fig.~\ref{#1}}

\usepackage{amsfonts}
\usepackage[applemac]{inputenc}
\usepackage[T1]{fontenc}
\usepackage{lmodern,textcomp}

\newcommand{\bb}[0]{\begin{eqnarray}}
\newcommand{\ee}[0]{\end{eqnarray}}

\def\(({\left(}
\def\)){\right)}                       
\def\e{{\rm e}}

\newcommand{\Tr}[1]{\mathrm{Tr}_{\{#1\}} }
\newcommand{\nn}{\nonumber}

\newcommand{\dd}{\mathop{}\mathopen{}\mathrm{d}}
\newcommand{\DD}{\mathop{}\mathopen{}\mathcal{D}}

\newcommand{\abs}[1]{|#1|}
\newcommand{\<}{\langle}
\renewcommand{\>}{\rangle}

\newcommand{\bara}[1]{\bar a_{#1}}
\newcommand{\barb}[1]{\bar b_{#1}}

\newcommand{\beq}{\begin{equation}}
\newcommand{\eeq}{\end{equation}}
\newcommand{\bea}{\begin{eqnarray}}
\newcommand{\eea}{\end{eqnarray}}

\newcommand{\ff}{\frac{1}{2}}
\newcommand{\ffi}{\frac{i}{2}}

\newcommand{\vc}[1]{{\bf c}_{#1}}

\newcommand{\prodd}[2]{\overrightarrow{\prod_{#1}^{#2}}}
\newcommand{\prodg}[2]{\overleftarrow{\prod_{#1}^{#2}}}

\newcommand{\Pf}{{\rm pf}}
\newcommand{\Perf}{{\rm perfect}}
\newcommand{\br}{{\bf r}}
\newcommand{\PQ}[2]{{Q}_{#1}(\{#2\})}

\usepackage{url} 
\usepackage {geometry}

\usepackage {listings}
\usepackage{lscape}

\usepackage {multicol}

\usepackage {setspace}

\makeatletter
\def\@makechapterhead#1{%
  \vspace*{50\p@}%
  {\parindent \z@ \raggedright \normalfont
    \interlinepenalty\@M
    \ifnum \c@secnumdepth >\m@ne
        \Huge\bfseries \thechapter\quad
    \fi
    \Huge \bfseries #1\par\nobreak
    \vskip 20\p@
  }}
\def\@makeschapterhead#1{%
  \vspace*{50\p@}%
  {\parindent \z@ \raggedright
    \normalfont
    \interlinepenalty\@M
    \Huge \bfseries  #1\par\nobreak
    \vskip 20\p@
  }}
\makeatother

\begin{document}

\begin{frontmatter}

\title{Exact solution of the $2d$ dimer model:\\ Corner free energy, correlation functions and combinatorics}

\author{Nicolas Allegra }
\address{Groupe de Physique 
Statistique, IJL, CNRS/UMR 7198, 
Université de Lorraine }
\fntext[myfootnote]{BP 70239, F-54506 Vand{\oe}uvre-lès-Nancy 
Cedex, France}
\ead[url]{nicolas.allegra@univ-lorraine.fr}

\begin{abstract}

In this work, some classical results of the pfaffian theory of the dimer model based on the work of Kasteleyn, Fisher and Temperley are introduced in a fermionic framework. Then we shall detail the bosonic formulation of the model {\it via} the so-called height mapping and the nature of boundary conditions is unravelled. The complete and detailed fermionic solution of the dimer model on the square lattice with an arbitrary number of monomers is presented, and finite size effect analysis is performed to study surface and corner effects, leading to the extrapolation of the central charge of the model. The solution allows for exact calculations of monomer and dimer correlation functions in the discrete level and the scaling behavior can be inferred in order to find the set of scaling dimensions and compare to the bosonic theory which predict particular features concerning corner behaviors. Finally, some combinatorial and numerical properties of partition functions with boundary monomers are discussed, proved and checked with enumeration algorithms.

\end{abstract}

\end{frontmatter}

\newpage

\tableofcontents

\section{Introduction}

Following \textsf{Onsager}'s solution of the $2d$ Ising model in the forties \cite{onsager1944crystal, kaufman1949crystalII,kaufman1949crystalIII}, the introduction of the \textsf{Bethe} ansatz \cite{bethe1931theorie} and the discovery of the machinery of transfer matrices, the field of exact solutions of lattice statistical physics models has exploded leading to the birth of a new domain of theoretical and mathematical physics known as {\it exactly solved models} \cite{baxter2007exactly}. The confluence of this new field with $2d$ conformal field theory (CFT) \cite{belavin1984infinite} discovered by \textsf{Belavin, Polyakov} and \textsf{Zamolodchikov} (see \cite{di1997conformal} for an extensive monography) had a huge impact in theoretical physics, from high energy to condensed matter, leading to a whole new level of understanding of classical and quantum integrable systems \cite{korepin1997quantum}. 

Initially, the dimer model (see \efig{figdimer}) has been introduced by physicists to describe absorption of diatomic molecules on a $2d$ subtrate \cite{fowler1937attempt}, yet it became quickly a general problem studied in various scientific communities.    
From the mathematical point of view, this problem known as perfect matching problem \cite{plummer1986matching}-- is a famous and active problem of combinatorics and graph theory \cite{flajolet2009analytic} with a large spectrum of applications. The enumeration of so-called \textsf{Kekulé} structures of molecular graphs in quantum chemistry are equivalent to the problem of enumeration of perfect matchings \cite{vukicevic2010applications,vukivcevic2011applications}. Besides, a recent connection between dimer models and $D$-brane gauge theories has been discovered \cite{hanany2005dimer,franco2006brane}, providing a very powerful computational tool.

The partition function of the $2d$ dimer model on the square lattice was solved independently using pfaffian methods \cite{kasteleyn1961statistics,fisher1961statistical,temperley1961dimer} for several boundary conditions, resulting in the exact calculation of correlation functions of two monomers along a row \cite{fisher1963statistical} or along a 
diagonal \cite{hartwig1966monomer,fisher2009toeplitz} in the scaling limit using \textsf{Toeplitz} determinants. For the general case of an arbitrary orientation, exact results are given in terms of the spin correlations of the $2d$ square lattice \textsf{Ising} model at the critical point \cite{perk84,kong87}. Other lattice geometries have been studied as well, {\it e.g.} the triangular lattice \cite{fendley2002classical}, the \textsf{Kagomé} lattice \cite{wang2007exact,wang2008dimers,wu2008dimers}, the triangular \textsf{Kagomé} lattice \cite{loh2008dimers}, the hexagonal lattice \cite{elser1984solution}, the star lattice \cite{fjaerestad2009classical}, or more complicated geometries \cite{yan2008dimer} (see \cite{wu2006dimers} for a review). The case of surface of high genius have been studied as well in \cite{costa2002dimers}.

The detailed study of the free energy and finite size effects began with the work of \textsf{Ferdinand} \cite{ferdinand1967statistical} few years after the exact solution, and has continued in a long series of articles using analytical \cite{izmailian05,izmailian2003exact,izmailian2006finite,izmailian2011dimer,nigro2012finite} and numerical methods \cite{kong2006monomer,kong2006packing,kong2006logarithmic,wu2011exact} for various geometries and boundary conditions. Some of these results have been motivated by the conformal interpretation of finite size effects \cite{blote1986conformal,cardy1988finite,cardy1989boundary} and leading to a somehow controversial result \cite{kong2006logarithmic,rasmussen2012refined} about the central charge of the underlying field theory.

Recent advances concern the analytic solution of the problem with a single monomer on the boundary of a $2d$ lattice \cite{2003dimers,wu06} thanks to a bijection du to \textsf{Temperley} \cite{temperley1974combinatorics}, boundary monomer correlation functions \cite{priezzhev2008boundary} and monomer localization phenomena \cite{bowick2007vacancy,jeng2008vacancy, poghosyan2011return}. Dimer models have regained interest because of its quantum version, the so-called quantum dimer model, originally introduced by \textsf{Rokhsar} and \textsf{Kivelson} \cite{rokhsar1988superconductivity} in a condensed matter context (see \cite{moessner2011quantum, fradkin2013field} for reviews) and equal to the classical model in a specific point of the parameter space. In this context, interactions between dimers have been considered at the classical and quantum level \cite{ alet2005interacting, alet2006classical, papanikolaou2007quantum,damle2012resonating}, leading to a richer phase diagram.

For the general monomer-dimer problem ({\it cf.} Appendix \ref{dimermonomer} for a definition) there is no exact solution except in $1d$, on the complete and locally tree-like graphs \cite{alberici2013solution} or scale free networks \cite{zhang2012monomer}. We can also mention that the matrix transfer method was used to express the partition function of the model  \cite{lieb1967,klein1990exact} and a very efficient method based on variational corner transfer matrix has been found by \textsf{Baxter} \cite{baxter1968dimers}, leading to  precise approximations of thermodynamic quantities of the model. From a more mathematical point of view, many results exist, one can mention the location of the zeros of the partition function \cite{heilmann1970monomers,heilmann1972theory}, series expansions of the partition function \cite{nagle1966new} and exact recursion relations \cite{ahrens1981paving}. Very recently a integrable version of the monomer-dimer model called monopole-dimer model has been proposed \cite{ayyer2013statistical}, sharing some qualitative features with the genuine monomer-dimer model. For $d>2$ lattices, no exact solution exists for the dimer model in general, but some analytical \cite{regge2000combinatorial,dhar2008exact} and numerical approaches \cite{bhattacharjee1983critical,huse2003coulomb} has been performed to study the phase diagram of the model. This lack of exact solution has been formalized in the context of computer science \cite{jerrum1987two}.

The prominence of the dimer model in theoretical physics and combinatorics also comes from the direct mapping between the square lattice \textsf{Ising} model without magnetic field and the dimer model on a decorated lattice 
\cite{mccoy1973two, kasteleyn1961statistics,fisher1961statistical,temperley1961dimer} and conversely from the mapping of the square lattice dimer model to a eight-vertex model \cite{baxter1972partition,wu1971ising}. Furthermore the magnetic field \textsf{Ising} model can be mapped onto the general monomer-dimer model \cite{heilmann1972theory}. Recently, some properties of the dimer model has been proved rigorously \cite{kenyon1997local,kenyon2000conformal,kenyon2001dominos} and various correlation functions has been studied as well \cite{ciucu2008dimer,ciucu2010emergence,dubedat2011dimers}. We could also mention the study of the double dimer model \cite{kenyon2011double,kenyon2014conformal} and the arctic circle phenomena \cite{elkies1992alternating,cohn1996local} in the Aztec geometry dimer model. 

\textsf{Grassmann} variables \cite{berezin}, thanks to their nilpotent properties, are very suitable to tackle combinatorial lattice models, and many of this models has been partially or entirely solved in the frame of fermionic field theory and the dimer model was one of them \cite{samuel1980use1,samuel1980use3}. In this context, we should mention the study of spanning trees and spanning forests \cite{caracciolo2007grassmann,caracciolo2004fermionic} as well as the edge-coloring problem \cite{fjaerestad20103, fjaerestad2012dimer} which is a special case of a more general loop model \cite{jacobsen1998field,kondev1998conformational}. The approach introduce presently has been developed by \textsf{Plechko} in a series of papers and has been widely used to solve in a very simple and elegant fashion many problems, {\it e.g.} the $2d$ \textsf{Ising} model \cite{plechko85a,plechko88}, boundary-field \textsf{Ising} model \cite{CF06}, the \textsf{Blume-Capel} model \cite{plechko2010fermions,clusel2008alternative} or more general spin models \cite{fortin2008second,clusel2009grassmann}. 

Recently a work based on this approach has been published extending the computation of the partition function of the dimer model with an arbitrary number of monomers \cite{allegra2014grassmannian}. In this present article, we continue the analysis of this solution focusing on the effect of surfaces and corners on the free energy and correlation functions. After reviewing the classical pfaffian theory and some of its most important features in a fermionic formulation, the bosonic version of the dimer model is introduced in order to compare CFT predictions with exact calculations. A special attention will be paid to the influence of boundary conditions to the expression of the free energy and critical exponents. At the end of the article, the exact expression of monomer correlations will be employed to bring out numerical identities about the partition function of the model with boundary monomers.

 \section*{Acknowledgement}
I am grateful to Jean-Yves Fortin for his early collaboration on this project and daily discussions. I would like to thank Jean-Marie Stéphan for crucial help and remarks on the height mapping and Christophe Chatelain, Jérôme Dubail, Malte Henkel, Jesper Jacobsen, Shahin Rouhani and Loïc Turban for insightful discussions and useful comments at different stages of the elaboration of this article. This work was partly supported by the Collège Doctoral Leipzig-Nancy-Coventry-Lviv (Statistical Physics of Complex Systems) of UFA-DFH.

\newpage

\section{Kasteleyn solution and bosonic formulation of the dimer model}\label{sect1}

In this section, the Kasteleyn-Fisher-Temperley theory of the dimer model is reminded in a fermionic field theory formulation and we show how this theory breaks down when monomers are introduced \cite{kasteleyn1961statistics,temperley1961dimer}. This theory was used to compute the partition functions, as well as dimer and monomer correlation functions in a perturbative way, leading to exact correlation exponents in the thermodynamic limit \cite{fisher1961statistical,fisher1963statistical}. At the end of the section, the bosonic formulation of the dimer model is presented {\it via} the so-called height mapping. The model will be then interpreted as a Coulomb gas theory of electric and magnetic charges. This theory will be very useful in order to compare with exact results about dimer and monomer correlations performed in the section \ref{sect3}. Special attention will be paid to boundary conditions and corner effects in a CFT framework, which will be crucial in order to interpret finite size effects to the free energy.

\subsection{Dimer model and nilpotent variables}

\begin{figure}[!h]
\center{
\includegraphics[scale=1]{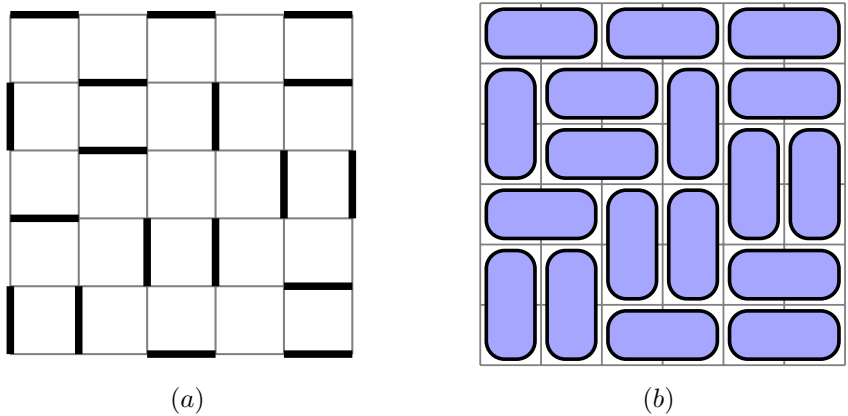}}
\caption{($a$) Perfect matching of the square lattice, and ($b$) its ''domino'' representation. This combinatorial problem reduces to the calculation of the partition function \eref{partitiondimer} with $t=1$.}
\label{figdimer}
\end{figure}

\begin{figure}[!h]
\center{
\includegraphics[scale=1.5]{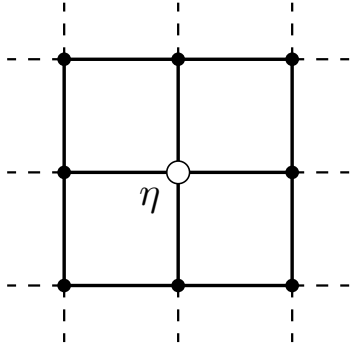}}
\caption{At every vertices, we put a nilpotent variable $\eta$, such that $\eta^{2}=0$ forbidden two dimers to occupy the same site.}
\label{figdimer0}
\end{figure}

\begin{figure}[!h]
\center{
\includegraphics[scale=1]{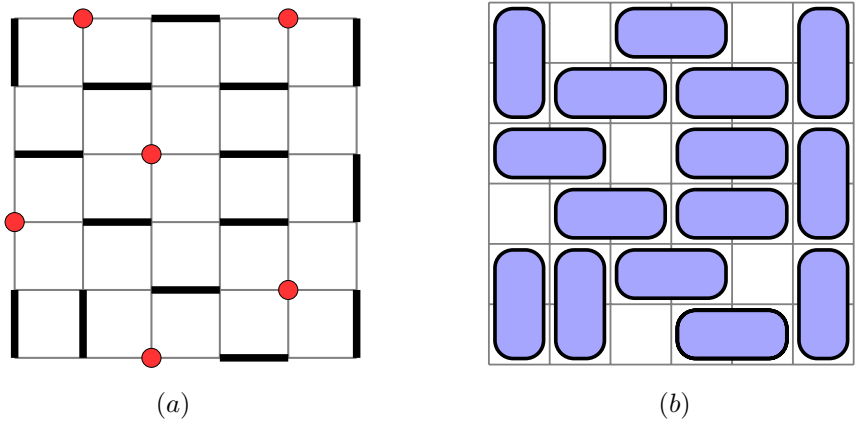}
}
\caption{($a$) Dimer model with 6 monomers, and ($b$) its ''domino'' representation.}
\label{figmonomer}
\end{figure}

A graph $\mathcal{G}$ is a pair of sets $(V, E)$, where $V$ is a finite set of vertices, and $E$ is a finite set of non-oriented edges.  We define the adjacent matrix (also called connectivity matrix) $A=(A_{ij})$, where the $ij$-entry is associated with the ordered pair of vertices $(v_{i},v_{i})$, then $A_{ij}=1$ if $v_{i}$ and $v_{j}$ are joined by an edge, and $0$ otherwise ({\it cf.}  \efig{figdimer} for the square lattice). The perfect matching number is the number of configuration with the property that each site of the lattice is paired with exactly one of its linked neighbors \cite{plummer1986matching}. In the language of theoretical physics, the enumeration of perfect matching of a planar graph $\mathcal{G}$ is equivalent to compute the partition function of the dimer model on the given lattice. In the simplest form, the number of dimers is the same in all the configurations, and the partition function is given by the 
equally-weighted average over all possible dimer configurations \footnote{Throughout this work, we will use the physics terminology and use the expresion {\it perfect matching} in some specific cases.}.
In the following, we will include equal fugacities $t$ for dimers, so that the average to be taken 
then includes weighting factors for dimers and we write the partition function as
\bb\label{partitiondimer}
Q_{0}[t]=\int\DD[\eta]\exp (-\beta\mathcal{H}),
\ee
where the Hamiltonian for the dimer written using commuting nilpotent variables (see Appendix \ref{grassmann}) can be written as
a sum over every vertices (see \efig{figdimer0}), preventing two dimers to occupy the same site
\bb\label{Hdimer}
\mathcal{H}=-\frac{t}{2}\sum_{ij}\eta_{i}A_{ij}\eta_{j},
\ee
where $A_{ij}$ is the adjacent matrix of the lattice considered. Let us put $\beta=1$ in the folllowing. The nilpotent variables can be seen as commuting Grassmann variables, or simply a product of two sets of standard Grassmann variables where $\eta_{i}=\theta_{i}\bar \theta_{i}$.
The perfect matching number of the graph $\mathcal{G}$ is equal to the partition function in the case $t=1$
\bb\nonumber
Q_{0}[1]&&=\int\DD[\eta]\exp \Big(\frac{1}{2}\sum_{ij}\eta_{i}A_{ij}\eta_{j}\Big)
=\int\DD[\theta,\bar \theta]\exp \Big(\frac{1}{2}\sum_{ij}\theta_{i}\bar\theta_{i}A_{ij}\theta_{j}\bar\theta_{j}\Big)\nonumber\\
&&={\rm hf} A.
\ee
We report the reader to the Appendix \ref{grassmann} for the definition of the haffnian of a matrix. In the second line we decomposed the nilpotent variables using two sets of Grassmann variables, and we finally found the well known graph theory result
\bb
\boxed{
\Perf{\ \mathcal{G}}={\rm hf} A.
}
\ee
Considering holes in the perfect matching problem is equivalent to remove rows and columns at the positions of the holes in the adjacent matrix (see \efig{figmonomer}). The resulting combinatorial problem is called the near-perfect matching problem. The partition function of the dimer model with a fixed number of holes (monomers) can be written as
\bb
Q_{n}[1]
&=&\int\DD[\theta,\bar\theta]\prod_{p=1}^{n} \theta_{q_{p}}\bar\theta_{q_{p}}\exp \Big(\frac{1}{2}\sum_{ij}\theta_{i}\bar\theta_{i}A_{ij}\theta_{j}\bar\theta_{j}\Big)\nonumber\\
&=&\int\DD[\theta,\bar\theta]\exp \Big(\frac{1}{2}\sum_{ij}\theta_{i}\bar\theta_{i}A^{\backslash \{q_{p}\}}_{ij}\theta_{j}\bar\theta_{j}\Big)\nonumber\\
&=&{\rm hf }A^{\backslash \{q_{p}\}},
\ee
where the index $n$ stands for the number of monomers in $Q_{n}[t]$. Finally, the result is the same but now the matrix $A^{\backslash \{q_{p}\}}$ is the adjacent matrix of the original graph with positions of the $n$ monomers removed.
Suppose we remove two sites $q_{1}$ and $q_{2}$ on the graph $\mathcal{G}$, then it is similar to introduce two nilpotent variables $\eta_{q_{1}}$ and $\eta_{q_{2}}$ on the lattice, the correlation function between these two monomers is then
\bb
\< \eta_{q_{1}}\eta_{q_{2}} \>=\<( \theta_{q_{1}}\bar\theta_{q_{1}})(\theta_{q_{2}}\bar\theta_{q_{2}})\>
&=&Q_{0}[1]^{-1}\int\DD[\eta]\exp \Big(\frac{1}{2}\sum_{ij}\theta_{i}\bar\theta_{i}A^{\backslash (q_{1},q_{2})}_{ij}\theta_{j}\bar\theta_{j}\Big)\nonumber\\
&=&{\rm hf} A^{\backslash (q_{1},q_{2})}{\rm hf}^{-1}A,
\ee
and more generally the $n$-point correlation function reads
\bb\label{permanent}
\Big\< \prod_{p=1}^{n}\eta_{q_{p}}\Big\>=\Big\< \prod_{p=1}^{n}\theta_{q_{p}}\bar\theta_{q_{p}}\Big\>
=\frac{{\rm hf} A^{\backslash\{q_{p}\}}}{{\rm hf} A} .
\ee
The partition function and correlations can be studied in the case $t\neq 1$ as well, in that case, the matrix elements of $A$ are $a_{ij}=\pm t$ and the generalization is straightforward. Generally correlations between monomers are equal to correlations between nilpotent variables in this framework, which can be written in terms of a ratio between two haffnian. Unlike the determinant which can be computed by a $\mathcal{O}(L^{3})$ time algorithm by Gauss elimination, there is no polynomial time algorithm for computing permanent. The problem of converting a permanent problem into a determinant problem is a long standing problem in pure mathematics, the simplest version of this problem, is called the P{\'o}lya permanent problem \cite{polya1913aufgabe}. Given a $(0,1)$-matrix $A:=(A_{ij})_{L\times L}$, can we find a matrix $B:=(B_{ij})_{L\times L}$ such that ${\rm perm} A={\rm det} B$ (or equivalently ${\rm hf} A=\Pf B$) where $B_{ij}=\pm A_{ij}$.

\subsection{Haffnian to Pfaffian conversion and Kasteleyn solution}

\begin{figure}[!h]
\center{
\includegraphics[scale=1]{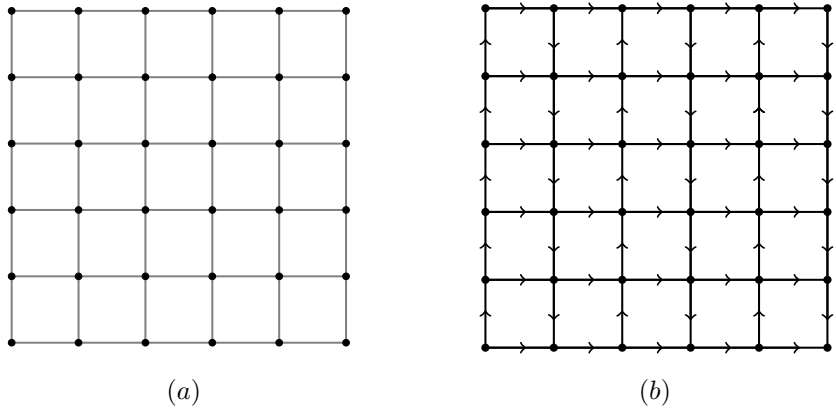}
}
\caption{($a$) Square lattice and ($b$) the orientation prescription of the Kasteleyn matrix.}
\label{figKasteleyn}
\end{figure}

The close-packed dimer model can be solved on any planar lattice by using Pfaffian techniques. These techniques were introduced in the early sixties by Kasteleyn \cite{kasteleyn1961statistics} and Temperley \cite{temperley1961dimer} and give a answer to the P{\'o}lya permanent problem for some particular conditions. The Kasteleyn theorem is a recipe to find a matrix\footnote{We will call this matrix the Kasteleyn matrix in the following.} $K$ in such way that ${\rm hf} A=\Pf K$, where the elements of the matrix are $K_{ij}=\pm1$. Kasteleyn theorem is based on a special disposition of arrows on the edges of a planar graph\footnote{This is a {\it condicio sine qua non} and the theorem is no longer valid for non-planar graph.}, the product of their orientation around any even-length closed path should be $-1$. Such a disposition is given in \efig{figKasteleyn}($b$) for the square lattice. We define an antisymmetric matrix ${K}$, where
\[
    {K}_{ij}=\left\{
                \begin{array}{ll}
                 1& \quad \text{if the arrow points from $i$ to  $j$}\\
                  -1& \quad \text{if the arrow points from $j$ to $i$}\\
                  0 & \quad \text{otherwise}
                \end{array}
              \right.
  \]
Kasteleyn theorem states that the perfect matching number of a given planar graph $\mathcal{G}$ is given by 
\bb
\Perf{\ \mathcal{G}}={\rm hf} A=\Pf{K},
\ee
which is equal to $\pm\sqrt{\det K}$. The $\pm$ sign is chosen to make the perfect matching number positive, henceforth we will omit this sign in the rest of the article. Differently, this pfaffian can be express in terms of Grassmann variables ({\it cf.} Appendix \ref{grassmann})
\bb
\Pf K=\int \DD[a] \exp\Big(\frac{1}{2}\sum_{i,j=1}^{N}a_{i}K_{ij}a_{j}   \Big).
\ee
For the square lattice, we can choose Boltzmann weights $t_{x}$ and $t_{y}$ for horizontal and vertical dimers, then the pfaffian can be computed by using Fourier transform and Kasteleyn found for free boundary conditions ({\it cf.} \cite{kasteleyn1961statistics} for details on calculations)
\bb\label{Kasteleyn}
\boxed{
Q_0[t_{x},t_{y}]=\prod_{p=1}^{M/2}\prod_{q=1}^{N/2}\left [ 4t_x^2\cos^2\frac{\pi 
p}{M+1}+4t_y^2\cos^2\frac{\pi q}{N+1} \right ].}
\ee
In table \ref{table1}, we compute $Q_{0}[1,1]$ using $\textsc{Mathematica}^{\circledR}$, for different $M$ and $N$ with $t_{x}=t_{y}=1$ (perfect matching number).
\begin{table}[h!]\center
\begin{tabular}{|c|cccccc|} \hline
$M$ $\backslash $ $N$ & ~$2$~ & ~$4$~   & ~$6$~  & ~$8$~ & ~$10$~& ~$12$~ \\ \hline
$2$   & $2$ & $5$  & $13$ & $34$                     & $89$    & $233$       \\
$4$   & $5$   & $36$  & $281$ & $2245$                     & $18061$   & $145601$         \\
$6$   & $13$   & $281$  & $6728$ & $167089$                     & $4213133$   & $106912793$        \\ 
$8$   & $34$   & $2245$  & $167089$ & $12988816$                     & $1031151241$    & $82741005829$        \\ 
$10$  & $89$   & $18061$  & $4213133$ & $1031151241$                     & $258584046368$   & $65743732590821$        \\ 
$12$  & $233$   & $145601$  & $106912793$ & $82741005829$                     & $65743732590821$    & $53060477521960000$       \\ \hline
\end{tabular}
\caption{Perfect matching number $Q_{0}[1,1]$ of the square lattice, for different $M$ and $N$.}
\label{table1}
\end{table}
All these values can be numerically checked using diverse algorithms which enumerate all the possible configurations on the square lattice (see \cite{krauth2006statistical} for details). In the rest of the paper we will omit the labels $t_{x}$ and $t_{y}$ in the partition function and just keep $Q_0$.

\subsubsection{Entropy in the thermodynamic limit}
The asymptotic form $L\rightarrow{\infty}$ of the partition function (for $M=N=L$) can be easily found from \eref{Kasteleyn}
\beq
Q_{0}\sim \exp{\frac{GL^{2}}{\pi}},
\eeq
where is the $G$ is Catalan constant\footnote{$ G=1^{-2}-3^{-2}+5^{-2}-7^{-2}+...=0.915 965 594...$}. The factor $G/\pi$ is the entropy by site of the dimer model on the square lattice. This entropy can also be calculated for other bipartite lattice like the honeycomb lattice, and for non-bipartite lattice like triangular, Kagomé lattice and triangular Kagomé lattice. The pfaffian method can be used to compute the partition function of the dimer model on various geometries and boundary conditions leading to different values of the entropy ({\it cf.} \cite{wu2006dimers} for review), as long as the lattice has a Kasteleyn orientation ({\it i.e.} planar according to the Kasteleyn theorem) and as long as $N\times M$ is even.  Obviously it is impossible to fill an odd size lattice with dimers, without leaving one site empty. We shall take notice later that the form of the free energy is strongly dependent of the parity of the lattice.

\subsubsection{Probabilities and correlations}

The Kasteleyn matrix is a very powerful objet, which gives us all the details about probabilities of presence of dimers \cite{fisher1963statistical}. For example the occupation probability $\mathbb{P}[i\rightarrow j]$ of a dimer on the link $ij$ is
\beq
\mathbb{P}[i\rightarrow j]={K}_{ij}\times {K}^{-1}_{ij},
\eeq
where $K_{ij}$ and $K_{ij}^{-1}$ are the corresponding matrix elements of the Kasteleyn matrix and its inverse. The probability of two dimers on the links $ij$ and $mn$ is 
\bb
\mathbb{P}[i\rightarrow j |  m\rightarrow n]=\det
\begin{pmatrix}
K_{ij}^{-1} & K_{im}^{-1}  \\
 K_{mj}^{-1} & K_{mn}^{-1}  \\
\end{pmatrix}.
\ee
In the rest of the paper, we shall use the term {\it correlation} even though this quantity is a normalized probability. Correlations between more than two dimers are available using the Kasteleyn matrix as well. It can be shown \cite{fisher1961statistical,fisher1963statistical} that dimer-dimer correlations on the square lattice decrease as the inverse square of the distance between the two dimers in the thermodynamic limit
\bb
\mathbb{P}(r)\sim r^{-2}.
\ee
Furthermore it has been also shown that correlations are always critical \cite{kasteleyn1963dimer} for bipartite lattices and {\it a contrario} exponential for non-bipartite lattices as the triangular lattice for example, where the fermionic theory underlying is a massive theory \cite{fendley2002classical}.

\subsection{Monomer correlation functions}

\begin{figure}[!h]
\center{
\includegraphics[scale=1]{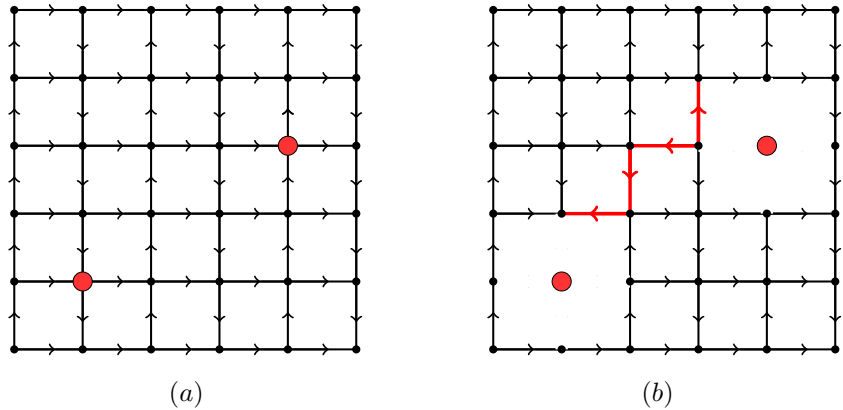}}
\caption{Modification of the Kasteleyn matrix in the presence of two monomers. The monomers (red dots) destroy the corresponding links and the orientation $(a)$ has to be changed to respect the proper orientation $(b)$.   }
\label{Kast1}
\end{figure}

Throughout this work the monomer-monomer correlation function $C$ will be defined as the ratio of the number of configurations with monomers at fixed positions to the number of configurations without monomers. Thus computing a monomer-monomer correlation is {\it stricto sensu} equivalent to compute the partition function with two sites (and all the links connected to these sites) deleted. Since such a graph is still planar, Kasteleyn’s construction is still applicable. The one complication is that we must ensure that on the new lattice with deleted sites, the number of arrows is still clockwise-odd. If all the monomers are located on the boundary of the lattice at ordinate $\{x_{i}\}$, there is no non-local defect lines between monomers (see \efig{Kast1}), and the modified  matrix $K^{\backslash\{x_{i}\}}$ defined from $K$ by removing all the rows and columns corresponding to the monomers positions has still the proper Kasteleyn orientation. Then the pfaffian of this modified Kasteleyn matrix  $K^{\backslash\{x_{i}\}}$ gives us the partition function of the dimer model with fixed monomer positions. It follows that the correlation function between two monomers on the boundary is
\bb\label{Kast-bound}
C(x_{1},x_{2}):=\frac{Q_{2}(x_{1},x_{2})}{Q_{0}}=\Pf\big( K^{-1}K^{\backslash(x_{1},x_{2})}\big)=\big\<a_{i}a_{j}\big\>,
\ee
where $x_{1}$ and $x_{2}$ are the positions of the two monomers. This pfaffian has been computed by Priezzhev and Ruelle ({\it cf.} \cite{priezzhev2008boundary} for details) in the thermodynamic limit for a arbitrary number of monomers at positions $\{x_{i}\}$, using a perturbative analysis of the matrix $K^{\backslash\{x_{i}\}}$ around the original Kasteleyn matrix $K$. The result for the $2n$-point correlation is given by
\bb\label{monomerbound}
C(x_{1},x_{2}...x_{2n})=\Pf \ C,
\ee
where the matrix element $C_{ij}:=C(x_{i},x_{j})$ is the 2-point function of a $1d$ complex free-fermion, equal to 
\bb
C_{ij}=-\frac{2}{\pi\abs{x_{i}-x_{j}}},
\ee
 if $x_{i}$ and $x_{j}$ are on opposite sublattices and $C_{ij}=0$ otherwise. For monomers in the bulk, the things are much more complicated. One sees that the product of arrows around a deleted site is now equal to $+1$ (see \efig{Kast1}).
We thus must construct a string of reversed arrows from one monomer to the second (see \efig{Kast1}($b$)). As long as the arrows are chosen to make all plaquette clockwise odd, the correlation is independent of the choice of the path. In the general case of bulk monomers, the relation \eref{Kast-bound} is no longer correct, because the matrix $K^{\backslash(x_{i},x_{j})}$ is no more a Kasteleyn matrix. Then correlations betweens two monomers defined by $Q_{2}(x_{i},x_{j})/Q_{0}$ is not equal to correlations between two Grassmann variables $\big\<a_{i}a_{j}\big\>$, but disorder operators must be add
\bb
\frac{Q_{2}(x_{i},x_{j})}{Q_{0}}=\Big\<a_{i}\exp{\Big(2\sum_{pq} K_{pq}a_{p} }a_{q}\Big)a_{j}\Big\>,
\ee
where the sum is over all the links connecting sites $i$ and $j$, to take account of the reversing line between the two monomers. Using a pfaffian perturbative analysis it was shown that monomer correlations decreases at the thermodynamic limit as \cite{fisher1961statistical,fisher1963statistical}
\bb
C(r)\sim r^{-1/2}.
\ee
This result is very similar to the construction of the spin correlation functions in the Ising model in terms of fermionic variables \cite{kadanoff1971determination,polyakov1987gauge}. In fact, on the square lattice, the monomer-monomer correlations was shown to have the same long-distance behavior as the spin-spin correlations in two decoupled Ising models, explained by the deep relation between the two correlation functions for the square lattice given by Perk and Au-Yang \cite{perk84}. These disorder operators are absent in the haffnian theory \eref{permanent}, and are the price to pay to solve the problem analytically.

\subsection{Mapping to a bosonic theory}

\subsubsection{Height mapping and Coulomb gas formalism}

\begin{figure}[!h]
\center{
\includegraphics[scale=1]{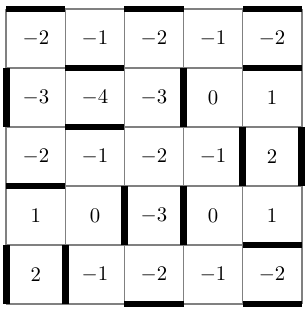}
\caption{Height mapping of the dimer model with free boundary conditions. For pedagogical purposes we will keep the field $h$ on the figures.}
\label{heightmap}}
\end{figure}

To any dimer covering we can associate a height on the dual lattice (on the plaquette) which is defined as follows \cite{zheng1989sine,levitov1990equivalence,ioffe1989superconductivity,henley1997relaxation}. When encircling an even vertex in the positive (counterclockwise) direction, the height $h$ increases by $+1$ upon crossing an empty edge and decreases by $-3$ upon crossing an edge that is covered by a dimer ({\it cf.} \efig{heightmap}). It is easy to notice that for the allowed configuration the average values $h_{\mathrm{vertex}}$ that the height variables can take at a given site of the direct lattice (a vertex) are $h_{\mathrm{vertex}}=\pm3/2$ or $h_{\mathrm{vertex}}=\pm1/2$. \footnote{We mention here that the height mapping remains valid for the interacting dimer model \cite{alet2005interacting, alet2006classical, papanikolaou2007quantum,damle2012resonating}, and it can be showed that, at the Kosterlitz-Thouless point, the interactions renormalize the free theory to another value of the stiffness $g=2/\pi$.}
On the other hand, a uniform shift of all the heights by one unit leads to an equivalent state. This mapping works {\it stricto sensu} for the close packed case. We will find it simpler to work with the rescaled height field $\phi=\frac{\pi}{2}h$.
By fixing the rescaled height at an arbitrary point, {\it e.g.} $\phi(0)=0$, these rules define the entire height function $\phi(\vec{r})$ uniquely. By integrating out the short distance fluctuations, one obtains an effective quadratic action $\mathcal{S}$ for the bulk height field $\phi(\vec{r})$, defined in the continuum, which corresponds to the long-wavelength modes
\bb\boxed{
\mathcal{S}[\phi]=\frac{g}{2} \int\dd x \dd y\ \big(\nabla \phi\big)^{2}.}
\ee
Here $g$ a constant which controls the stiffness of the height model. It is {\it a priori} unknown. The field has to be  invariant under the transformation $\phi=\phi+2\pi n$ to respect lattice symmetries. The derivation of this gaussian field has been actually done rigorously by mathematicians \cite{kenyon2000conformal,kenyon2001dominos}. Electric charges $e$ correspond to vertex operators appearing in the Fourier expansion of any operator periodic in the height field. Dual magnetic charges $m$ correspond to a dislocation in the height field and correspond to the dual vertex operator
\bb\nonumber
V_{e}(z)&=&:\e^{ie\phi}:\\
V_{m}(z)&=&:\e^{i m\psi}:
\ee
where $\psi$ is the dual field of $\phi$ and defined as
\bb
\partial_{i}\psi=\epsilon_{ij}\partial_{j}\phi.
\ee
These operators are primary operators of the $c=1$ CFT. The scaling dimension associated to the insertion of a particle with electromagnetic charge $(e,m)$ is given by 
\bb
x_{g}(e,m)=\frac{e^{2}}{4 \pi g}+\pi gm^{2}.
\ee
For example, two monomers on opposite sublattices correspond to two charges $m=1$ and $m=-1$. It is known from exact results \cite{fisher1961statistical,fisher1963statistical} that, the exponent for bulk monomer-monomer correlations is $1/2$, it fixes $g_{\rm{free}}=1/4\pi$ for the stiffness constant of the gaussian field theory describing the free dimer model. We saw previously that bulk dimer-dimer exponent is $2$. Hence the bulk monomer and dimer scaling dimensions defined by  $x^{(m)}_{b}:=x_{\frac{1}{4\pi}}(0, 1)$ and $x^{(d)}_{b}:=x_{\frac{1}{4\pi}}(1, 0)$ are
\bb
{\rm free \ dimer:} \ \ g_{\rm{free}}=\frac{1}{4\pi} \rightarrow \left\{
    \begin{array}{ll}
        x^{(d)}_{b}:=x_{\frac{1}{4\pi}}(1, 0)=1 \\
        x^{(m)}_{b}:=x_{\frac{1}{4\pi}}(1, 0)=1/4.
    \end{array}
\right.
\ee
In this theory the conformal spin of an operator is defined by $s(e,m)=em$, then monomers and dimers are spinless particles but fermions which are order-disorder composite operators \cite{kadanoff1971determination} have magnetic and electric charges and carry spins $1/2$. The fermion operator has then scaling dimension $x_{\frac{1}{4\pi}}(1/2,1)=1/2$. It is also possible to define parafermion operators which obey fractional statistics \cite{fradkin1980disorder} for particular values of the stiffness $g$. The use of such a mapping to study correlation functions dates back to Bl{\"o}te, Hilhorst, and Nienhuis \cite{nienhuis1984triangular,blote1982roughening}. The neutral 2-point correlation functions for vertex operators are
 then given by the standard formula \cite{Nienhuis1987}
\bb
&&\Big\<V_{m=+1}(z)V_{m=-1}(0)\Big\>\sim z^{-2x_{\frac{1}{4\pi}}(0,1)}=z^{-1/2}\nn\\
&&\Big\<V_{e=+1}(z)V_{e=-1}(0)\Big\> \sim z^{-2x_{\frac{1}{4\pi}}(1,0)}=z^{-2}.
\ee
The general monomer $2n$-point function is given by the product 
\bb
\Big\<V_{m_{1}=\pm1}(z_{1})...V_{m_{2n}=\pm1}(z_{2n})\Big\>\sim\frac{ \prod_{i<j} (b_{i}-b_{j})^{1/2} \prod_{k<l} (w_{k}-w_{l})^{1/2}   }{\prod_{p<q} (b_{p}-w_{q})^{1/2} },
\ee
where $\{b_{k}\}$ and $\{w_{k}\}$ are the sets of the even/odd sublattice coordinates. In this Coulomb gas interpretation of the dimer model, monomers located on the same sublattice are seen as repealing equal charges, and monomers on opposite sublattice are seen as attractive opposite charges. It is known from exact results \cite{priezzhev2008boundary,allegra2014grassmannian} that, the exponent for monomer correlations is $1$ in the case (see \eref{monomerbound}) where the monomers are restricted to live on a boundary.  Hence the surface monomer scaling dimension is $x^{(m)}_{s}:=1/2$.

\subsubsection{Rectangular geometry and boundary conditions}

\begin{figure}[!h]
\center{
\includegraphics[scale=0.9]{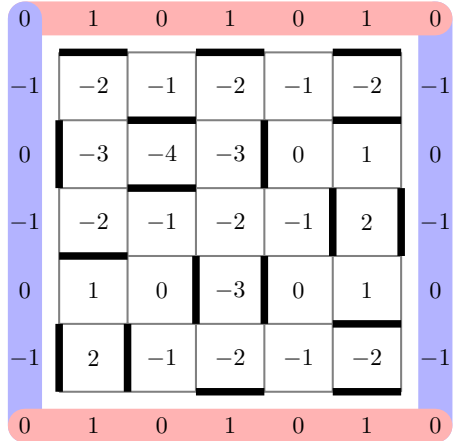}
\caption{Height mapping of the dimer model with free boundary conditions. The different values of the height $h$ on each sides of each corner are induced by a {\it bcc} operator of scaling dimension $1/32$. The dimension does not change if one chooses a rectangle $M\neq N$.}
\label{heightmap1}}
\end{figure}

On \efig{heightmap1}, we show a configuration of dimers on the square lattice with free boundary conditions. The choice of these conditions impose very specific boundary conditions for the height variable. Indeed, we observe that the value of the height $h$ is $(..101..)$ along the boundary until the corner and $(..-10-1..)$ along the next boundary \cite{stephan2014emptiness} (a similar analysis has been done for the strip geometry \cite{stephan2011phase,stephan2012renyi}).
Then we have, from the point a view of the field theory two different averaged boundary conditions either sides of the corner, let us call the boundary fields $  h_{b}=1/2$ and ${\bar h_{b}}=-1/2$. The proper way to understand how corners change the behavior of the height field is through boundary CFT ({\it cf.} \cite{henkel1999conformal} for introduction). Hence it is natural to introduce local operators \cite{cardy1986effect} acting on the corners of the domain, these objets called {\it boundary condition changing} operators ({\it bcc}) can be showed to be primary operators of the CFT \cite{cardy1989boundary} . A careful analysis shows that the difference of boundary conditions has non negligible consequences in the thermodynamic limit and the precise procedure based on the contribution of theses operators on the free energy can be found in \cite{stephan2011phase}. It has been shown that the dimension of the {\it bcc} operator which creates a field shift of value $\Delta\phi_{b}=\frac{\pi}{2}(h_{b}-{\bar h_{b}})=\phi_{b}-{\bar \phi_{b}}$ on the corner is 
\bb\label{heightshift}
h_{{\it bcc}}= \frac{g}{2\pi}\Delta\phi_{b}^{2},
\ee
with $g_{\rm{free}}=1/4\pi$ here. In the previous situation without any monomer, the corner shift in the height $h$ is equal to $\Delta h_{b}=1$ then $\Delta\phi_{b}=\pi/2$, hence the dimension of the corner {\it bcc} operator is $h_{{\it bcc}}=1/32$. Furthermore, the addition of a monomer on the boundary or at the corner will change the value of the field, and we will behold further that it will be relevant in order to study quantities as the free energy and correlation functions. A general framework to study partition functions and conformal boundary states on the rectangular geometry with different boundary conditions in a boundary CFT framework has been developed recently \cite{bondesan2012conformal,bondesan2013rectangular}. In the section \ref{sect3}, finite size effects to the free energy for the free and interacting cases will be study in this CFT framework, and the influence of these ${\it bcc}$ operators will be crucial to identify the correct underlying central charge of the theory. Furthermore the presence of monomers on the boundaries or at the corners will change again the values of the {\it bcc} operators which will be important for the study of surface and corner correlation functions.

\section{Exact partition function and corner free energy}\label{sect2}

In this section, the fully-detailed Grassmann solution of the dimer model with an arbitrary number of monomers is presented, which will lead to the exact form of correlation functions in a pfaffian formulation. In particular, we show that the problem become simpler when we consider boundary monomers, and a closed expression for correlations can be found. Hereinafter the solution of the dimer model on a odd size lattice with one boundary monomer is introduced, in agreement with the Tzeng-Wu solution \cite{2003dimers}. Those solutions will be used to extract finite-size scaling behaviors of the free energy in a CFT framework, where {\it boundary changing conditions} operators has to be carefully study to infer the central charge of the model, in contradiction with a recent article \cite{izmailian2014exact}.

\subsection{Plechko pfaffian solution}
Here we just recall the framework of the Plechko solution \cite{plechko93,plechko94} of the dimer model. As we have seen in the section \ref{sect1}, the partition function can be written for a general graph using nilpotent variables
\bb
Q_0=\int\DD[\eta] \exp\Big(\frac{1}{2}\sum_{i,j=1}^{N}\eta_{i}A_{ij}\eta_{j}\Big).
\ee
We are now working on the square lattice with free boundary conditions, then the partition function reads
\bb
Q_{0}=\int \DD[\eta]  \prod_{m,n}^{L}(1+t_x\eta_{mn}\eta_{m+1n})(1+t_y\eta_{mn}\eta_{mn+1}),
\ee
where $t_{x}$ and $t_{y}$ are the horizontal and vertical Boltzmann weight, and $m$ and $n$ refer to the coordinates. The partition function can be written using Grassmann variables (see Appendix \ref{appendixplechko} for details), this leads to a block representation of the action in the momentum space, for momenta inside the 
reduced sector $1\le p,q\le L/2$. The four components of these vectors will 
be written $c_{\alpha}^{\mu}$ with $\mu=1\cdots 4$, leading to
\bb\label{Q0int}
{Q}_0=\int \DD[c]\exp {S}_0[c],
\ee
with $S_0[c]=\ffi c_{\alpha}^{\mu}M_{\alpha}^{\mu\nu}c_{\alpha}^{\nu}$\footnote{Repeated indices are implicitly summed over.}, where 
the antisymmetric matrix $M$ is defined by
\bb\label{matrixM}
M_{\alpha}=
\begin{pmatrix}
0 & 0 & a_y(q) & a_x(p)  \\
 0 & 0 & -a_x(p) & a_y(q) \\
-a_y(q) & a_x(p) & 0 & 0 \\
-a_x(p) & -a_y(q) & 0 & 0 \\
\end{pmatrix}
\ee
with 
\bb\nonumber
a_x(p)&=&2t_x\cos\frac{\pi p}{L+1},\\
a_y(q)&=&2t_y\cos\frac{\pi q}{L+1}.
\ee
This matrix can be written as 
\bb
M_{\alpha}&=& a_x(p)\Gamma_x+a_y(q)\Gamma_y,
\ee
where the matrices $\Gamma_x$ and $\Gamma_y$ are
\bb
\Gamma_x&=&
\begin{pmatrix}
0 & 0 & 0 & 1  \\
 0 & 0 & -1 & 0 \\
0 & 1 & 0 & 0 \\
-1 & 0 & 0 & 0 \\
\end{pmatrix},\;
\Gamma_y=
\begin{pmatrix}
0 & 0 & 1 & 0  \\
 0 & 0 & 0 & 1 \\
-1 & 0 & 0 & 0 \\
0 & -1 & 0 & 0 \\
\end{pmatrix},
\ee
with $\Gamma_x^2=\Gamma_y^2=-1$. Hence the expression \eref{Q0int} is directly related to the pfaffian of the matrix $M$ ({\it cf.} \ref{grassmann} for details) and is simply equal to
\bb
Q_0&=&\prod_{\alpha} \Pf M_{\alpha}\\
&=&\prod_{p,q}^{L/2}[a_x(p)^2+a_y(q)^2].\nn
\ee
Finally one simply obtains the following well known result
\beq
Q_0=\prod_{p,q=1}^{L/2}\left [ 4t_x^2\cos^2\frac{\pi p}{L+1}+4t_y^2\cos^2\frac{\pi q}{L+1} \right ].
\eeq
The fermionization can also be performed for toroidal boundary conditions. We refer here to the experience with the $2d$ Ising model on a torus \cite{plechko85a}. The final result can be written in terms of a combination of the
periodic-antiperiodic boundary conditions for fermions $c_{M+1n=\pm c_{1n}}$ and $c_{mN+1=\pm c_{m1}}$.

\subsection{Pfaffian solution with $2n$ monomers.}

\subsubsection{General case}

\begin{figure}[h!]
\begin{center}
\includegraphics[scale=1]{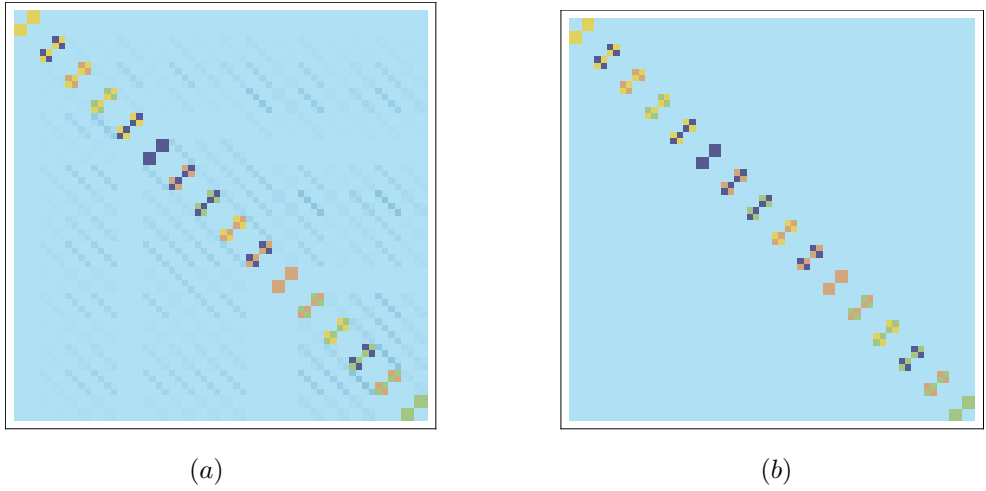}
\caption{($a$) Graphical representation of the matrix $W=M+V$. ($b$) When the monomers are on the boundaries of the domain, the off diagonal term $V$ vanishes and the matrix $W$ reduces to the matrix $M$. }\label{matrixW}
\end{center}
\end{figure}

Let us now consider the case where an even\footnote{The case with an odd number of monomers can be studied as well as we shall see later.} number of monomers are present on the lattice at different positions $\br_i=(m_i,n_i)$ with $i=1,\cdots,2n$. The partition function $\PQ{2n}{\br_i}$ is the number of all possible configurations
with the constraint imposed by the fixed monomers. This quantity can be evaluated by inserting nilpotent variables $\eta_{m_in_i}$ in the partition function, which prevents possible dimers to occupy sites $\br_i$. It can be useful to introduce an additional Grassmann variable $h_i$ such that $\eta_{m_in_i}=\int \dd h_i\exp(h_i\eta_{m_in_i})$.
These insertions are performed at point $\br_i$ in $Q_{0}$, and the integration over $\eta_{m_in_i}$
modifies $L_{m_in_i}\rightarrow L_{m_in_i}+h_i$. However, by moving the $\dd h_i$ variable to the left of the remaining ordered product, a minus sign is introduced in front of each $\bar b_{mn_i-1}$ in $\bar B_{mn_i}$ for all $m>m_i$. We can replace $\bar b_{mn-1}$ by $\epsilon_{mn}\bar b_{mn-1}$ such that $\epsilon_{mn_i}=-1$ for $m>m_i$, and $\epsilon_{mn}=1$ otherwise. The integration is then performed on the remaining variables $(a,\bar a,b,\bar b)$ as usual, so that $\PQ{2n}{\br_i}$ 
can be expressed as a Gaussian form, with a sum of terms corresponding to the monomer insertion, resulting to the following form for the the partition function
\begin{figure}[h!]
\center{
\includegraphics[scale=1.1]{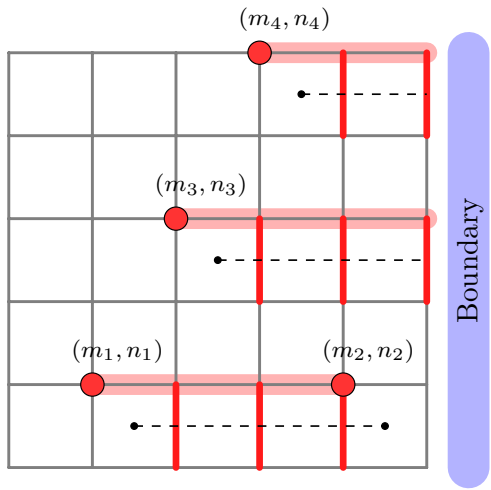}}
\caption{Typical dimer configuration for a $6\times 6$ square lattice. The 
dashed line is the extra term that arises from moving towards the 
border the Grassmann field conjugated to the monomer. This is 
equivalent to change the sign of the $t_{y}$ couplings (red links) 
from the monomers to the boundary. When two monomers are located 
on the same horizontal line, the change of sign concerns only 
the couplings between the two monomers.}\label{fig_2D}
\end{figure}
\bb\label{hole_action}
\boxed{
\PQ{2n}{\br_i}
=\int\DD[c]\DD[h] \exp\left [ {S}_0+\sum_{\br_i}c_{m_in_i}h_i+2t_y\sum_{\br_i,m=m_i+1}^L(-1)^{m+1}c_{mn_i-1}c_{
mn_i}\right ].}
\ee
where $\int\DD[c]\DD[c]=\int\prod_{mn} \dd c_{mn}\prod_{i}\dd h_{i}$ where the index $i$ runs over the set of positions $\br_i=(m_i,n_i)$ of the $2n$ monomers.
The inclusion or monomers is equivalent to inserting a magnetic field $h_i$ at points $\br_i$, as well as a sum of quadratic terms $c_{mn_i-1}c_{mn_i}$ running from the hole position to the boundary on the right (see \efig{fig_2D}). Another possibility would be to join two monomers by a line of terms by moving $\dd h_{m_in_i}$ until $\dd 
h_{m_jn_j}$ as in the Kasteleyn theory. In this case, the additional quadratic terms in the action starting from $\br_i$ and 
ending on the boundary have to be treated in the computation of the Grassmannian
integral. We first rewrite $S_0$ in the Fourier space using the block partition
label $\alpha=(p,q)$ for momenta $p$ and $q$ inside the reduced sector 
$1\cdots L/2$, and vectors $\vc{\alpha}=^t(c_{pq},c_{-pq},c_{p-q},c_{-p-q})$. Also the 4 components of vector $\vc{\alpha}$ will be written $c_{\alpha}^{\mu}$ where $\mu=1\cdots 4$. Then $S_0=\frac{i}{2} c_{\alpha}^{\mu}M_{\alpha}^{\mu\nu}c_{\alpha}^{\nu}$, where the antisymmetric quadratic form $M_{\alpha}$ is defined by \eref{matrixM}.  The part of the field interaction can be 
Fourier transform as before with a linear field $H_{pq}$ depending on $h_i$s and we obtain
\bb
\sum_{\br_i}c_{m_i,n_i}h_i=\sum_{p,q=1}^Lc_{pq}H_{pq}=\sum_{\alpha,\mu}c_{\alpha
}^{\mu}H_{\alpha}^{\mu}.
\ee
The last contribution connecting the monomers to the boundary can be written as $\frac{i}{2}c_{\alpha}^{\mu}V_{\alpha\beta}^{\mu\nu}c_{\beta}^{\nu}$, with matrix $V_{\alpha\beta}=V_{pq,p'q'}$ given by
\bb\nn
V_{pq,p'q'}=
\sum_{\br_i}\frac{8t_y(-1)^{n_i}}{(L+1)^2}
\left \{
\sum_{m=m_i+1}^L\sin\frac{\pi 
p m}{L+1}\sin\frac{\pi p' m}{L+1}
\right \}
\left [
\sin\frac{\pi q (n_i-1)}{L+1}\sin\frac{\pi q' n_i}{L+1}
-\sin\frac{\pi q' (n_i-1)}{L+1}\sin\frac{\pi q n_i}{L+1}
\right ].
\ee
The different components $V_{\alpha\beta}^{\mu\nu}$ are given implicitly, for the first elements, by
$V_{\alpha\beta}^{11}=V_{pq,p'q'}$, $V_{\alpha\beta}^{12}=V_{pq,-p'q'}$,
$V_{\alpha\beta}^{21}=V_{-pq,p'q'}$, and so on. Then the total fermionic action
contains three terms 
\bb
S=\frac{i}{2} 
c_{\alpha}^{\mu}M_{\alpha}^{\mu\nu}c_{\alpha}^{\nu}+\frac{i}{2} 
c_{\alpha}^{\mu}V_{\alpha\beta}^{\mu\nu}c_{\beta}^{\nu}
+c_{\alpha}^{\mu}H_{\alpha}^{\mu}.
\ee
The first two terms contain only modes of the same sector $\alpha$, and the last connects modes 
from different sectors $\alpha$ and $\beta$. Matrices $M_{\alpha}^{\mu\nu}$ and $V_{\alpha\beta}^{\mu\nu}$ are
antisymmetric: $M_{\alpha}^{\mu\nu}=-M_{\alpha}^{\nu\mu}$ and 
$V_{\alpha\beta}^{\mu\nu}=-V_{\beta\alpha}^{\nu\mu}$. Also $V_{\alpha\alpha}=0$. Then the quantity $\PQ{2n}{\br_i}$ can be formally written as
\bb\label{action_bulk}
\PQ{2n}{\br_i}
&=&\int\DD[c]\DD[h]\exp\left (\ffi
c_{\alpha}^{\mu}
[M_{\alpha}^{\mu\nu}\delta_{\alpha\beta}
+V_{\alpha\beta}^{\mu\nu}
]
c_{\beta}^{\nu}
+c_{\alpha}^{\mu}H_{\alpha}^{\mu}
\right )\nn\\
&=&
\int\DD[c]\DD[h]\exp\left (\ffi 
c_{\alpha}^{\mu}W_{\alpha\beta}^{\mu\nu}
c_{\beta}^{\nu}
+c_{\alpha}^{\mu}H_{\alpha}^{\mu}
\right ),
\ee
with $W_{\alpha\beta}^{\mu\nu}=\delta_{\alpha\beta}M_{\alpha}^{\mu\nu}
+V_{\alpha\beta}^{\mu\nu}$ ({\it cf.} \efig{matrixW}($a$)). By construction, $W$ is antisymmetric and 
satisfies $W_{\alpha\beta}^{\mu\nu}=-W_{\beta\alpha}^{\nu\mu}$.  
This matrix can be represented as a block matrix of global size $L^2\times L^2$
\bb
W=\begin{pmatrix}
M_{\alpha=(1,1)} & V_{(1,1),(1,2)} & V_{(1,1),(1,3)} \cdots &  \\
\multicolumn{1}{c}{$\upbracefill$}&\\
\multicolumn{1}{c}{\scriptstyle 4 \times 4 \;{\rm matrix}}&\\
 -V_{(1,1),(1,2)} & V_{(1,2)} & V_{(1,2),(1,3)} & \cdots \\
-V_{(1,1),(1,3)} & -V_{(1,2),(1,3)} & M_{(1,3)} & \cdots  \\
\vdots & & & \\
\multicolumn{3}{c}{$\upbracefill$}\\
\multicolumn{3}{c}{\scriptstyle L^2/4 \;{\rm blocks}}\\
\noalign{\vspace{-\normalbaselineskip}}
\end{pmatrix}
\vspace{\normalbaselineskip}
\ee
where each of the $(L^2/4)\times(L^2/4)$ blocks is a $4\times 4$ matrix. Labels 
$\alpha$ are ordered with increasing momentum $(1,1),(1,2)\cdots (1,L/2),(2,1)\cdots$ In 
the action the linear terms in $c_{\alpha}^{\mu}$ can be removed using a linear change of variables $c_{\alpha}^{\mu}\rightarrow 
c_{\alpha}^{\mu}+g_{\alpha}^{\mu}$, where $g_{\alpha}^{\mu}$ are Grassmann constants. After some algebra, we find the correct values for the constants that eliminate the linear contribution are given by $g_{\alpha}^{\mu}=i(W^{-1})_{\alpha\beta}^{\mu\nu}H_{\beta}^{\nu}$. After substitution of these values in the overall integral, and a rescaling of variables $c_{\alpha}\rightarrow c_{\alpha}/\sqrt{i}$, the action becomes a product over variables $c$ and $h_i$
\bb\nn
\PQ{2n}{\br_i}
&=&
\int\DD[c]\DD[h]\exp\left [\ff 
c_{\alpha}^{\mu}W_{\alpha\beta}^{\mu\nu}
c_{\beta}^{\nu}
-\ffi(W^{-1})_{\alpha\beta}^{\mu\nu}H_{\alpha}^{\mu}H_{\beta}^{\nu}
\right ]
\\ \nn
&=&\Pf(W)\int\DD[h]\exp\left [-\ffi(W^{-1})_{\alpha\beta}^{\mu\nu}H_{\alpha}^{\mu}H_{\beta}^{\nu}
\right ].
\ee
The fields $H_{\alpha}^{\mu}$ can be expressed with $h_i$ as 
$H_{\alpha}^{\mu}=\sum_{i=1}^{2n}\Lambda_{i,\alpha}^{\mu}h_i$,
where coefficients $\Lambda_{i,\alpha}^{\mu}$ can be rewritten using 
a 4-dimensional vector, such as
\bb
{\bf \Lambda_{i,\alpha}}=\frac{2}{L+1}\sin\frac{\pi 
p m_i}{L+1}\sin\frac{\pi q n_i}{L+1}{\bf \Lambda_i}=:r_i(\alpha){\bf 
\Lambda_i},
\ee
where the vector
\bb
{\bf \Lambda_i}=
\begin{pmatrix}
  i^{m_i+n_i}  \\
-i^{-m_i+n_i} \\ -i^{m_i-n_i} \\ i^{-m_i-n_i}
\end{pmatrix}
\vspace{\normalbaselineskip},
\ee
depends only on the location parity of the monomer $\br_i$ in the bulk.
Functions $r_i(\alpha)$ are normalized 
$\sum_{\alpha}r_i(\alpha)r_j(\alpha)=\delta_{ij}$.
Then we obtain the following formal and compact expression for 
$\PQ{2n}{\br_i}$
\bb\nn
\PQ{2n}{\br_i}
&=&\Pf(W)\int\DD[h]\exp\left (\ff
\sum_{i,j}h_ih_j 
\Lambda_{i,\alpha}^{\mu}(W^{-1})_{\alpha\beta}^{\mu\nu}
\Lambda_{j,\beta}^{\nu}
\right )
\\ \label{action_bulk_res}
&=:&\Pf(W)\int\DD[h]\exp\left (\ff
\sum_{i,j}h_ih_j C_{ij}
\right ).
\ee
Finally, we found a pfaffian expression of the partition function
\bb
\boxed{
\PQ{2n}{\br_i}=\Pf(W)\Pf(C).
}\ee
We verify easily that the matrix $C$ is antisymmetric by using the antisymmetry property of $W$ or $W^{-1}$
\bb\nn
C_{ji}&=&\Lambda_{j,\alpha}^{\mu}(W^{-1})_{\alpha\beta}^{\mu\nu}
\Lambda_{i,\beta}^{\nu}
=\Lambda_{j,\beta}^{\nu}(W^{-1})_{\beta\alpha}^{\nu\mu}
\Lambda_{i,\alpha}^{\mu}\\
&=&-\Lambda_{j,\beta}^{\nu}(W^{-1})_{\alpha\beta}^{\mu\nu}
\Lambda_{i,\alpha}^{\mu}\nn\\
&=&-C_{ij}.
\ee
Then, $C_{ij}$ can be formally expressed as a scalar product 
$C_{ij}=\sum_{\alpha,\beta}\left \< {\bf \Lambda_{i,\alpha}}| 
W^{-1}_{\alpha,\beta}|{\bf \Lambda_{j,\beta}}\right \>$. $Q_{2n}$ is therefore a product of two pfaffians where the 
monomer locations are specified in matrix $W$. 
The matrix $V$ can be rewritten using additional matrices after considering the different components 
$(\mu,\nu)$. We can indeed express $V$ using four functions 
$u_k^{a,s}(\alpha,\beta)$, and 
$v_k^{a,s}(\alpha,\beta)$, for each monomer at location $\br_k=(m_k,n_k)$, with 
$m_k<L$, and such that
\bb
V_{\alpha\beta}&=&-2t_y
\sum_{\br_k}\sum_{c,c'}
u_k^c(\alpha,\beta)\Gamma_{cc'}v_k^{c'}(\alpha,\beta),
\ee
where the $\Gamma_{cc'}$ are defined as
\bb\nn
\Gamma_{sa}&=&
\begin{pmatrix}
1 & 0 & 0 & 0  \\
 0 & 1 & 0 & 0 \\
0 & 0 & -1 & 0 \\
0 & 0 & 0 & -1 \\
\end{pmatrix},
\Gamma_{aa}=
\begin{pmatrix}
0 & 1 & 0 & 0  \\
1 & 0 & 0 & 0 \\
0 & 0 & 0 & -1 \\
0 & 0 & -1 & 0 \\
\end{pmatrix},
\Gamma_{ss}=
\begin{pmatrix}
0 & 0 & 1 & 0  \\
 0 & 0 & 0 & 1 \\
-1 & 0 & 0 & 0 \\
0 & -1 & 0 & 0 \\
\end{pmatrix},
\Gamma_{as}=
\begin{pmatrix}
0 & 0 & 0 & 1  \\
0 & 0 & 1 & 0 \\
0 & -1 & 0 & 0 \\
-1 & 0 & 0 & 0 \\
\end{pmatrix}.
\ee
Functions $u$ and $v$ are given by
\bb\nn
u_k^s(\alpha,\beta)&=&\frac{2}{L+1}\sum_{m=m_k+1}^L\sin\frac{\pi 
p m}{L+1}\sin\frac{\pi p' m}{L+1},\\ \nn
u_k^a(\alpha,\beta)&=&\frac{2}{L+1}\sum_{m=m_k+1}^L(-1)^{m+1}\sin\frac{\pi 
p m}{L+1}\sin\frac{\pi p' m}{L+1},\\ \nn
v_k^s(\alpha,\beta)&=&\frac{2}{L+1}\left [
\sin\frac{\pi q n_k}{L+1}\sin\frac{\pi 
q'(n_k-1)}{L+1}
+\sin\frac{\pi q'n_k}{L+1}\sin\frac{\pi q(n_k-1)}{L+1}
\right ]
,\\ 
v_k^a(\alpha,\beta)&=&\frac{2(-1)^{n_k}}{L+1}\left [
\sin\frac{\pi q n_k}{L+1}\sin\frac{\pi 
q'(n_k-1)}{L+1}
-\sin\frac{\pi q'n_k}{L+1}\sin\frac{\pi q(n_k-1)}{L+1}
\right ].
\ee
More explicitly the result is
\bb\nn
u_k^s(\alpha,\beta)&=&
\frac{\sin\frac{\pi(p-p')(L+1/2)}{L+1}-
\sin\frac{\pi(p-p')(m_k+1/2)}{L+1}}{2(L+1)\sin\frac{\pi(p-p')}{2(L+1)}}-\frac{\sin\frac{\pi(p+p')(L+1/2)}{L+1}-
\sin\frac{\pi(p+p')(m_k+1/2)}{L+1}}{2(L+1)\sin\frac{\pi(p+p')}{2(L+1)}},
\\ \nn
u_k^a(\alpha,\beta)&=&-\frac{\cos\frac{\pi(p-p')(L+1/2)}{L+1}-(-1)^{m_k}
\cos\frac{\pi(p-p')(m_k+1/2)}{L+1}}{2(L+1)\cos\frac{\pi(p-p')}{2(L+1)}}+\frac{\cos\frac{\pi(p+p')(L+1/2)}{L+1}-(-1)^{m_k}
\cos\frac{\pi(p+p')(m_k+1/2)}{L+1}}{2(L+1)\cos\frac{\pi(p+p')}{2(L+1)}},
\\ \nn
v_k^s(\alpha,\beta)&=&\frac{2}{L+1}\left [
\sin\frac{\pi q n_k}{L+1}\sin\frac{\pi 
q'(n_k-1)}{L+1}
+\sin\frac{\pi q'n_k}{L+1}\sin\frac{\pi q(n_k-1)}{L+1}
\right ],
\\ 
v_k^a(\alpha,\beta)&=&\frac{2(-1)^{n_k}}{L+1}\left [
\sin\frac{\pi q n_k}{L+1}\sin\frac{\pi 
q'(n_k-1)}{L+1}
-\sin\frac{\pi q'n_k}{L+1}\sin\frac{\pi q(n_k-1)}{L+1}
\right ].
\ee
This close solution to the dimer model with an arbitrary number of monomers at fixed location can be formally used to get access to some informations about the general monomer-dimer model (see Appendix \ref{dimermonomer}). 

\subsubsection{Boundary monomers}

\begin{figure}[h!]
\begin{center}
\includegraphics[scale=1.1]{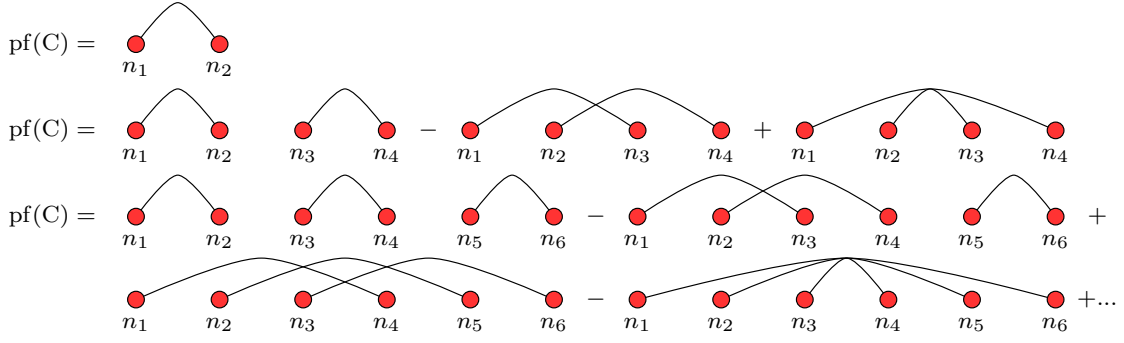}.
\caption{Diagrammatical representation of the pfaffian of $C$ for $2$, $4$ and $6$ monomers}
\label{default1}
\end{center}
\end{figure}

Although the general problem remains in principle tractable, we can simplify it further by considering monomers on the boundary $m_i=L$ of the rectangle only, 
with the convention $n_i>n_j$ if $i>j$. 
\begin{figure}[h!]
\center{
\includegraphics[scale=1.2]{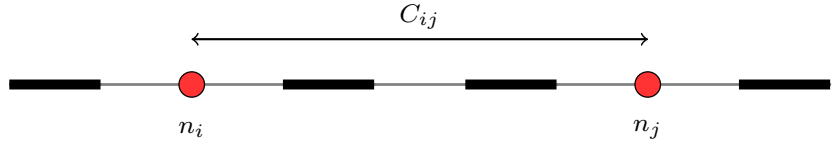}}
\caption{Configuration of 2 monomers on the boundary at the position $n_{i}$ and $n_{j}$ correlated by $C_{ij}$. This correlation decreases as the inverse of the distance and vanishes if $n_i$ and $n_j$ have the same parity.}\label{boundmono}
\end{figure}
When the monomers are on the boundaries of the domain, the off diagonal term $V$ vanishes and the matrix $W$ reduces to the matrix $M$ ({\it cf.} \efig{matrixW}($b$)). In that case, the previous action contains no defect line term, only the couplings with fields remain and the problem can be transformed into a $1d$ system of particles on a chain (see \efig{boundmono}). Using the previous Fourier transform for the $c$'s variables, the magnetic field term becomes
\bb\nn
& &\sum_{\br_i}c_{m_i,n_i}h_i=\sum_{p,q=1}^Lc_{pq}H_{pq},
\\ \label{Hpq}
H_{pq}&=&\sum_{\br_i} \frac{2i^{m_i+n_i}}{L+1}\sin\frac{\pi pm_i}{L+1} \sin\frac{\pi qn_i}{L+1}h_i.
\ee
Then the partition function can be integrated on the $c_{mn}$'s using the block 
partition $(p,q)$ and we find 
that 
\bb
\PQ{2n}{\br_i}={Q}_0\int \dd h_1\cdots \dd h_{2n}\exp {S}_H,
\ee
where
\bb\nn
{S}_H&=&\sum_{p,q=1}^{L/2}\frac{it_x\cos\frac{\pi p}{L+1}}{2t_x^2\cos^2\frac{\pi 
p}{L+1}+2t_y^2\cos^2\frac{\pi q}{L+1}}
\left (
H_{p-q}H_{-pq}+H_{pq}H_{-p-q}
\right )
\\ \nn
&+&\frac{it_y\cos\frac{\pi q}{L+1}}{2t_x^2\cos^2\frac{\pi 
p}{L+1}+2t_y^2\cos^2\frac{\pi q}{L+1}}
\left (
H_{-pq}H_{-p-q}+H_{pq}H_{p-q}
\right ).
\ee
Grassmann fields $H_{pq}$ has the following properties: $H_{-pq}=-H_{pq}$ and $H_{-p-q}=-H_{p-q}$.
In that case, the first sum on the right hand side of the previous equation 
vanishes due to the anti-commuting
property, and we can reduce the field-dependent action to one single sum
\bb\label{SH}
{S}_H=\sum_{p,q=1}'
\frac{it_y\cos\frac{\pi q}{L+1}}{t_x^2\cos^2\frac{\pi 
p}{L+1}+t_y^2\cos^2\frac{\pi q}{L+1}}
H_{pq}H_{p-q},
\ee
where the prime symbol is meant for summation over half of the modes 
$p,q=1\cdots L/2$. This action would vanish if all the $n_i$ were for example even, since the Grassmann
fields satisfy in this case $H_{pq}H_{p-q}=-H_{pq}^2=0$. In general, the field 
action can be rewritten as a quadratic form over the real-space fields $h_i$: 
${S}_H=\sum_{i<j}C_{ij}h_ih_j$, where the elements of the  
matrix $C$ are antisymmetric $C_{ij}=-C_{ji}$, and equal to
\bb\label{CijBound}
\boxed{
C_{ij}=\frac{4\left [ (-1)^{n_i}-(-1)^{n_j} \right ]}{(L+1)^2}\sum_{p,q=1}^{L/2}
\frac{i^{1+n_i+n_j}t_y\cos\frac{\pi q}{L+1}\sin^2\frac{\pi p}{L+1}}
{t_x^2\cos^2\frac{\pi p}{L+1}+t_y^2\cos^2\frac{\pi q}{L+1}}
\sin\frac{\pi q n_i}{L+1}\sin\frac{\pi q n_j}{L+1}
}
\ee
These elements are zero if $n_i$ and $n_j$ have the same parity and in general 
the integration over the field 
variables $h_i$ leads directly to a pfaffian form for the partition function 
$\PQ{2n}{\br_i}={Q}_0(-1)^n\Pf(C)$.
The $(-1)^n$ factor comes from the rearrangement of the measure $\dd h_1\cdots 
\dd h_{2n}=(-1)^n\dd h_{2n}\cdots \dd h_1$, so that the $2n$-function reads
\bb
\int \prod_{i=1}^{2n}\dd h_{i}\exp\left (-\sum_{i<j}C_{ij}h_ih_j\right )=\Pf C.
\ee
This sign could also be absorbed in the definition of matrix elements 
$C_{ij}\rightarrow -C_{ij}$. The resulting partition function is always positive with this 
definition. For example, if there are 2 monomers at the boundary, $\Pf(\hat 
C)=-C_{12}$, and 4 monomers\footnote{As a straightforward application, the number of perfect matching with a monomer on each corner can be computed and the result is $Q^{\rm{corner}}_{4}=\{1, 8, 784, 913952, 1211936774...\}$} leads to $\Pf(\hat 
C)=C_{12}C_{34}-C_{13}C_{24}+C_{14}C_{23}$ ({\it cf.} \efig{default1}). 
We should notice that the expression \eref{CijBound} is an exact closed expression for the 2-point correlation between monomers on the boundary $m_{i}=m_{j}=L$, leading to an explicit result for the partition function with $2,4...2n$ monomers. 
We will show in a next section how to find the correlation between monomers on different (opposite or adjacent) boundaries of the lattice. 

\subsubsection{Single monomer on the boundary}

\begin{figure}[h!]
\center{
\includegraphics[scale=1.3]{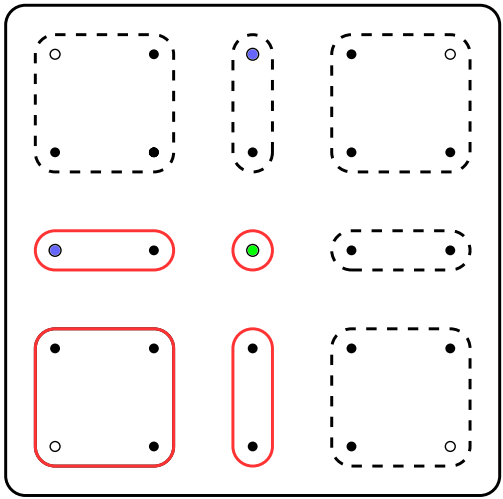}
}
\caption{Block partition of the Fourier modes for $L$ odd (here $L=5$). There 
are 3 blocks of momenta $(p,q)$ taken into account into the integration, plus 
one point (green dot) at location $(\ff(L+1),\ff(L+1))$. The first block of 
momenta $(p,q)$ is represented in the red square for values $1\le p,q \le \ff(L-1)$, with 
corresponding momenta $(\pm p,\pm q)$ (open dots). Then there are 2 additional 
lines of values $p=\ff(L-1)$ with $1\le q \le \ff(L-1)/2$, and corresponding momentum $(\ff(L-1),-q)$, and $q=\ff(L-1)$ with $1\le p \le \ff(L-1)/2$, and corresponding momentum $(-p,\ff(L-1))$, blue dots.}
\label{block_odd}
\end{figure}

We can recover the partition function for one monomer on the boundary using the 
previous analysis. In this case the size $L$ has to be odd in order to 
accommodate for the presence of one single monomer. The action \eref{Q0int} is 
still valid, but the Fourier transform leads to a
different block arrangement for the bulk terms \eref{S0} which are represented 
by the red zones in \efig{block_odd}
\bb\label{S0odd}\nn
S_0&=&2it_x\sum_{p,q\ge 1}^{\ff(L-1)}\cos\frac{\pi p}{L+1}\left 
(c_{pq}c_{-p-q}+c_{p-q}c_{-pq}\right )+2it_x\sum_{p\ge 1}^{\ff(L-1)}\cos\frac{\pi 
p}{L+1}c_{p\ff(L+1)}c_{-p\ff(L+1)}
\\ \nn
&+&2it_y\sum_{p,q\ge 1}^{\ff(L-1)}\cos\frac{\pi q}{L+1}\left (c_{pq}c_{p-q}+c_{-pq}c_{-p-q}\right )+2it_y\sum_{q\ge 1}^{\ff(L-1)}\cos\frac{\pi 
q}{L+1}c_{\ff(L+1)q}c_{\ff(L+1)-q}.
\ee
$S_0$ contains Fourier modes that cover the Brillouin zone (see 
\efig{block_odd}) except for term $c_{\ff(L+1)\ff(L+1)}$ which is located in the 
middle of the zone and not present in the sums \eref{S0odd}. The integration 
over Grassmann variables $c_{pq}$ is therefore zero in absence of
coupling with this mode. Inserting a single monomer on the boundary at 
location $\br=(L,n)$ is equivalent, as previously demonstrated, to inserting a 
Grassmann field $h$ and a field contribution $S_H=c_{Ln}h$ in the 
action, which can be expanded using Fourier transformation
\bb\nn
S_{H}&=&\frac{2i^{L+n}}{L+1}\sum_{p,q}(-1)^{p+1}\sin\frac{\pi p}{L+1} 
\sin\frac{\pi qn}{L+1}c_{pq}h
\\ &=&\sum_{p,q}c_{pq}H_{pq}.
\ee
Since there is only one Grassmann field, all terms $H_{pq}$ 
are only proportional to $h$, and therefore the quadratic form in \eref{SH} is 
zero. However one term contributes to the integration over 
$c_{\ff(L+1)\ff(L+1)}$,
and corresponds to $c_{\ff(L+1)\ff(L+1)}H_{\ff(L+1)\ff(L+1)}$. 
After integration over the remaining $c_{pq}$ and $h$ variables, the 
partition function can then be factorized as
\bb\nn
Q_1&=& \frac{-2i^{n+1}\sin(\pi n/2)}{L+1}\prod_{p,q=1}^{\ff(L-1)}\left [ 4t_x^2\cos^2\frac{\pi 
p}{L+1}+4t_y^2\cos^2\frac{\pi q}{L+1} \right ]\prod_{p=1}^{\ff(L-1)}
2t_x\cos\frac{\pi p}{L+1}
\prod_{q=1}^{\ff(L-1)}
2t_y\cos\frac{\pi q}{L+1}.
\ee
This result is consistent to the fact that a monomer can be put only at 
odd site locations. We can use the formula 
$\prod_{p=1}^{\ff(L-1)}2\cos(\frac{\pi 
p}{L+1})=\sqrt{\frac{L+1}{2}}$, to simplify the previous expression and recover 
the Tzeng-Wu \cite{2003dimers} solution 
\bb\label{tzeng}
\boxed{
Q_1=\prod_{p,q=1}^{\ff(L-1)}\left [ 4t_x^2\cos^2\frac{\pi 
p}{L+1}+4t_y^2\cos^2\frac{\pi q}{L+1} \right ]\times
(t_xt_y)^{\ff(L-1)}\times [-i^{n+1}\sin(\pi n/2)].
}
\ee
We can notice that the partition function with one monomer in a system of size $L\times L$ ($L$ odd) is equal to the partition function without monomers on a lattice of size $L-1\times L-1$. The probability is therefore constant for all location of the monomer, at even sites only, the last term in bracket being equal to zero ($n$ odd) or unity ($n$ even), proving that the monomer is fully delocalized on the boundary, unlike the bulk case where monomers are actually localized in a finite region of the domain \cite{bowick2007vacancy,jeng2008vacancy,poghosyan2011return}.

\subsection{Corner free energy and the central charge controversy}

The study of finite size effects in statistical physics is a long standing and still active field of research \cite{privman1990finite}. {\it A fortiori} the possibility to solve a model in a non homogeneous geometry \cite{igloi1993inhomogeneous} is of prime interest for the understanding of behavior of physical systems in real situations. In the case of the dimer model on the rectangle with free boundary conditions, the system admit surfaces and corners, both of them play an important role in the behavior of the free energy in the thermodynamic limit. The exact solution of the close packing dimer model \eref{Kasteleyn} on a even lattice ($M\times N=2p$) allows for the study of the finite size effect of the free energy, and the finite size analysis has already been performed in the early time of the dimer model history \cite{ferdinand1967statistical}. 

Furthermore the exact solution \eref{tzeng} of the dimer model on a odd lattice ($M\times N=2p+1$) with a monomer on a boundary allows for the study of the free energy in that case as well. Adding a finite number of monomers in the dimer model is equivalent to a zero density of monomers in the continuum limit. Hence the presence of monomers does not give any contribution in the expression of the free energy, the only feature which plays a crucial role is the parity of the size of the lattice (even or odd). Because they are the simplest expressions of a even (odd) lattice with an even (odd) number of monomers, these two partition functions are sufficient to study all the details of the asymptotic limit of the free energy. In the following let us choose the square geometry $M=N=L$ for simplicity\footnote{The entire procedure can be extended to the case $M\neq N$ where the aspect ratio has to be taken into account, and no significant change appears \cite{kleban1991free}.}. The free energy on a finite lattice of typical length $L$ at criticality has the generic following form \footnote{The presentation here closely follows \cite{stephan2013logarithmic}.} 
\bb\label{FSC}
\mathcal{F}=L^{2}f_{\rm{bulk}}+ Lf_{\rm{surface}}+f_{0}+a \log L+o\Big(\frac{\log L}{L}\Big).
\ee
The first term is the extensive contribution of the free energy, whereas the second one represents contribution from the lattice surface. In general, the coefficients $f_{\rm{bulk}}$ and $f_{\rm{surface}}$ are non-universal, but the coefficient $f_{0}$ is assumed to be universal, depending only on the shape and boundary conditions of the system.
Universal properties of critical models appear in the subleading corrections and the value of $f_{0}$ is known to be simply related to the central charge $c$ of the underlying conformal field theory. The study of statistical systems and their field theory representation in the presence of corners has been covered extensively, {\it e.g.} Ising and Potts model, loop model and percolation \cite{jacobsen2010bulk,vernier2012corner,kovacs2012corner,kovacs2014corner}, using various theoretical and numerical machinery. In two dimensions, as pointed out by Cardy and Peschel  \cite{cardy1988finite}, the universal contribution to the free energy of a critical system in a domain with a corner with angle $\theta$ has been determined using the complex transformation 
\bb
z\rightarrow z^{\theta/\pi}
\ee
which maps the upper half-plane onto the corner and looking at the holomorphic component of the stress-energy tensor in the corner. This mapping gives us the explicit form of the logarithmic contribution $\mathcal{F}_{\rm{corner}}=a \log L$ in \eref{FSC} and the result is $\frac{c}{24}\big(\frac{\theta}{\pi}-\frac{\pi}{\theta}\big)\log L$. 
\begin{figure}[h!]
\center{
\includegraphics[scale=1.5]{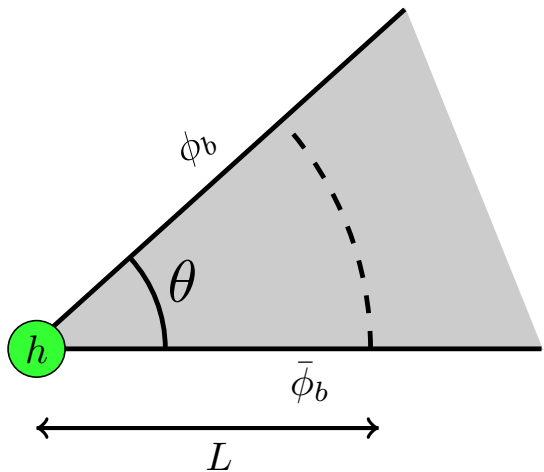}
}
\caption{Corner of angle $\theta$ in a system of typical length $L$, in our situation $\theta=\pi/2$. The green dot symbolize a {\it bcc} operator of scaling dimension $h_{bcc}$ which changes the value of the field either sides of the corner.}
\label{corner1}
\end{figure}
It turns out that a additional complication arises because of the {\it bcc} operators \cite{cardy1989boundary} acting in the corners ({\it cf.} \efig{corner1}) as we saw in the section \ref{sect1}. In that particular case, the Cardy-Peschel contribution is slightly modified to taking into account this change of boundary conditions \cite{kleban1992domain}, and the logarithmic corner contribution becomes
\bb\label{cornerF}
\boxed{
\mathcal{F}_{\rm{corner}}=\Big[\frac{\pi}{\theta}h_{bcc}+\frac{c}{24}\Big(\frac{\theta}{\pi}-\frac{\pi}{\theta}\Big)  \Big]\log L
}
\ee
where $h_{bcc}$ is the scaling dimension of the {\it bcc} operator. This {\it bcc} operator changes the boundary condition either sides of the corner in the height field representation. In their recent paper about corner free energy contribution in the free boundary conditions dimer model with one monomer at the boundary \cite{izmailian2014exact}, the authors analyzed the asymptotic contribution of the four corners of the rectangular system without taking into account this {\it bcc} operator, ie taking $h_{bcc}=0$ in the previous formula, and concluded that the central charge of the dimer model is $c=-2$. It is beyond the scope of this paper to enter into details in the wide area of conformal analysis of finite size effects (see \cite{stephan2013logarithmic} for more details), nor all the literature of $c=-2$ models, but just to pointing out the difference of result when one looks carefully at the {\it bcc} operator contribution in the corner free energy. In their paper, the authors found that the contribution of the four corners of the $L\times L$ ($L$ odd) lattice is \footnote{{\it cf.} \cite{izmailian2014exact} for details of the asymptotic calculation.}
\bb\label{coin1}
\mathcal{F}_{\rm{corner}}=\frac{1}{2}\log L.
\ee 
The CFT formula \eref{cornerF} gives in the square geometry case ($\theta=\pi/2$)
\bb\label{coin2}
\mathcal{F}_{\rm{corner}}=&\Big[\frac{-c}{4}+2\big(h^{(1)}_{bcc}+h^{(2)}_{bcc}+h^{(3)}_{bcc}+h^{(4)}_{bcc}\big)\Big]\log L,
\ee
where $h^{(\nu=1..4)}_{bcc}$ are the dimensions of the four {\it bcc} operators living on the corners. Taking $h_{bcc}=0$ leads  {\it de facto} to $c=-2$, suggesting that the dimer model may be a logarithmic CFT (LCFT) \cite{izmailian05, priezzhev2008boundary,rasmussen2012refined}. This statement is also based on the mapping of the dimer model to the spanning tree model \cite{temperley1974combinatorics} and, equivalently, to the Abelian sandpile model \cite{majumdar1992equivalence} which both belong to a $c=-2$ LCFT, facts which we do not dispute here. The problem with this analysis is the oversight
 of the {\it bcc} operators acting on the corners. We know that the partition function with one monomer on the boundary does not depend of the location of the monomer, then let us choose to put it on the corner for simplicity.
\begin{figure}[!h]
\center{
\includegraphics[scale=1.1]{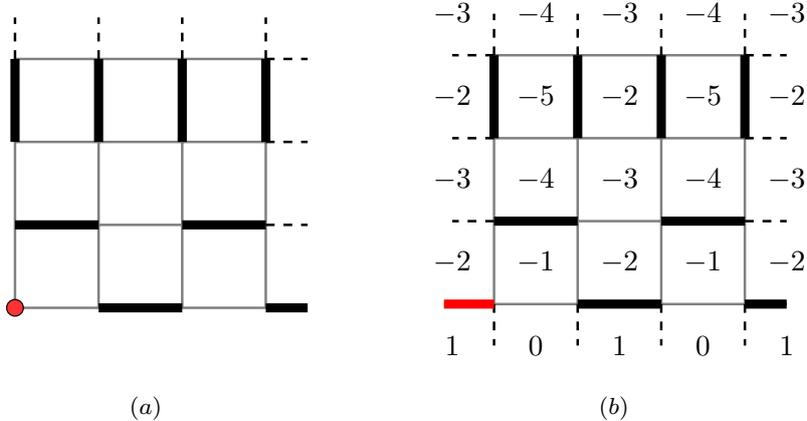}
\caption{($a$) Representation of the corner region of a lattice with one monomer at the corner. ($b$) The monomer can be represented by a virtual red dimer which gives the same configurations. The different values of the height field either sides of the corner are induced by a {\it bcc} operator of scaling dimension $9/32$.}
\label{heightmap2}
}
\end{figure}%
In this following case, the height field is shifted either sides of the corner, precisely we obtain the value $h=(..0101..)$ in one side and $h=(..-2-3-2-3..)$ on the other side (see \efig{heightmap2}), making a height shift of $\Delta\phi_{b}=3\pi/2$, therefore using \eref{heightshift}, the dimension of the {\it bcc} operator on the corner is $9/32$ (which is actually the dimension of the original {\it bcc} plus the dimension of the corner monomer operator as we will see in the next section). The three other corners induce a height shift of $\pi/2$, thus the dimension of the {\it bcc} operators is $1/32$ as in the case where monomer are absent. Finally we find
\bb
\mathcal{F}_{\rm{corner}}=&\Big[\frac{-c}{4}+2\Big(\frac{9}{32}+\frac{1}{32}+\frac{1}{32}+\frac{1}{32}\Big)\Big]\log L.
\ee
The comparison with the asymptotic result \eref{coin1} gives us the value of the central charge of the free bosonic field theory, {\it i.e.} $c=1$\footnote{One can notice that the same result holds if the monomer is somewhere on the boundary but not at the corner, in that case, there is five {\it bcc} opeartors, four on each corners of dimension $1/32$ and another responsible for the shift at the surface with dimension $1/2$ then using \eref{cornerF}
\bb
\mathcal{F}_{\rm{corner}}=&\Big[\frac{-c}{4}+\frac{1}{2}+2\Big(\frac{1}{32}+\frac{1}{32}+\frac{1}{32}+\frac{1}{32}\Big)\Big]\log L=\frac{1}{2}\log L,
\ee
in accordance with the fact that the partition function with one boundary monomer does not depend of its location. 
}. Obviously this result seems completely natural in the height mapping framework or in the free complex fermion representation of the dimer model \eref{hole_action}, nonetheless the presence of  {\it bcc} operators acting on the corners has never been extensively studied in this context, leading to misinterpretation of asymptotic results and thus to a different value of the central charge of the theory. In the pure dimer situation \eref{Kasteleyn}, the four {\it bcc} operators has dimension $1/32$, we claim that the corner free energy should be equal to
\bb
\mathcal{F}_{\rm{corner}}&=&\Big[\frac{-1}{4}+2\big(4h_{{\it bcc}}\big)\Big]\log L=0.
\ee
This result nicely agrees with the literature, where no corner free energy term has never been found in the free boundary conditions close packed dimer model, neither theoretically \cite{brankov1995isomorphism} nor numerically \cite{kong2006logarithmic}, strengthening our analysis\footnote{The same analysis can be done for the same model with periodic boundary conditions in one direction and free in the other, leading to the same conclusion about the central charge.}. Previously another type of finite size effect analysis \cite{izmailian05} has been performed for the close packing dimer model on a strip with periodic and free boundary conditions, and it has been shown that the result depends strongly on the parity of the length of the strip as it should be, and the notion of effective central charge has to be introduce \cite{itzykson1986conformal}. Though the comparison with the CFT result gives the value $c=-2$, we claim here that the analysis of the free energy in the height mapping formulation of boundary conditions, leads again to $c=1$. In the following section, we shall compute analytically correlation functions between monomers where boundary fields has to be properly interpreted and we will show that the exact solution fully agrees with this theory, reinforcing the present result about the significance of these {\it bcc} operators. Finally we conclude, that the free bosonic formulation of the dimer model allows for the complete study and interpretation of finite size effects in a CFT context, and unified all the results known about this very simple but not trivial model.

\section{Exact correlations: discrete and continuous cases}\label{sect3}

In this section, detailed computations of correlations are performed in terms of disorder operators. A particular attention will be paid to the special case of boundary monomers, where a closed-form expression is obtained, valid for any of the four boundaries of the rectangle. Then, numerical evaluations of the exact pfaffian solution are done and diverse monomer and dimer correlations are obtained for bulk, surface and corner cases, leading to the whole set of scaling dimensions which can be compared to the bosonic theory with $g=1/4\pi$. 

\subsection{Fermion correlations and disorder operators}

The addition of monomers in the dimer model is therefore equivalent to inserting a magnetic field $h_i$ at points $\br_i$, as well as a line of defect running from the monomer position to the right boundary $m=L$. If two monomers have the same ordinate $n_i=n_j$, the line of defects will only run between the two monomers and will not reach the boundary. This can be viewed as an operator acting on the links crossed by the line and running from a point on the dual lattice to the boundary 
on the right-hand side. More specifically, we can express the correlation 
functions, after integration over the fermionic magnetic fields $h_i$, as an average over composite fields
\bb
\frac{\PQ{2n}{\br_i}}{Q_{0}}&=&\Big\<  \prod_{\{\br_{i}\}} c_{m_in_i} 
\exp \Big(2t_{y}\sum_{m=m_{i}+1}^{L}(-1)^{m+1}c_{mn_{i-1}}c_{mn_{i}} \Big)  
\Big\>_{0}\nn\\
&=&\Big\<  \prod_{\{\br_{i}\}} c_{m_in_i} \mu(\br_{i}+{\bf 
e_4})\Big\>_{0}\nn\\
&=&\Big\<  \prod_{\{\br_{i}\}} \Psi_4(\br_{i})\Big\>_{0},
\ee
where $\mu(\br+\bf{e}_4)$ is a fermionic disorder 
operator, whose role is to change the sign of the vertical links across its 
path starting from vector $\br+\bf{e}_4$ on the dual lattice ({\it cf.} \cite{allegra2014grassmannian} for details). 
 The integration $\<\cdots\>_0$ is performed 
relatively to the 
action $S_0$. 
Likewise the Kasteleyn theory, where disorder lines are absent on the boundary and where correlations between monomers correspond to correlations between Grassmann variables \eref{Kast-bound}, here the correlation between monomers on the boundaries are exactly correlation functions between the fermionic fields
\bb
\frac{\PQ{2n}{\br_i}}{Q_{0}}&=&\Big\<  \prod_{\{\br_{i}\}} c_{m_in_i} 
\Big\>_{0}.
\ee
This result about monomer correlations written in terms of disorder operators is the fermionized version of the Coulomb gas framework, where monomers act like dual magnetic charges which create a dislocation of the height field and correspond to the vertex operator of the corresponding bosonic field theory.

\subsection{Perturbative expansion of the 2-point function}

In the case where the monomers are on the boundaries, we were able to compute exactly the 2-point correlation function \eref{CijBound} in the discrete case. In the bulk case, the things are much more complicated and an exact closed-form expression on the discrete level seems out of reach. Nevertheless, a perturbative expansion can be performed to evaluate the pfaffian expression of the correlation function.
We start from the exact pfaffian expression of matrix $C$
\bb
C_{ij}&=&\Lambda_{i,\alpha}^{\mu}(W^{-1})_{\alpha\beta}^{\mu\nu}
\Lambda_{j,\beta}^{\nu}.
\ee
The inverse matrix $W^{-1}$ can be computed using formally the expansion 
\bb\label{Wexp}
W^{-1}=(M+V)^{-1}=M^{-1}-M^{-1}V M^{-1}+\mathcal{O}(V^{2}).
\ee
 In particular it is convenient to write the inverse matrix $M^{-1}$ as
\bb
M^{-1}_{\alpha}&=&\bar a_x(p)\Gamma_x+\bar a_y(q)\Gamma_y,
\ee
with 
\bb
&&\bar a_x(p)=-\frac{a_x(p)}{a_x(p)^2+a_y(q)^2}\nn\\
&&\bar a_y(q)=-\frac{a_y(q)}{[a_x(p)^2+a_y(q)^2}.
\ee
In the following we will consider only the first of this expansion
\bb\label{Cij}\nn
C_{ij}&=&\sum_{\alpha,\beta}\left \< {\bf \Lambda_{i,\alpha}}| 
W^{-1}_{\alpha\beta}|{\bf \Lambda_{j,\beta}}\right \>
=\sum_{\alpha}\sum_{\mu}
r_i(\alpha)\bar a_{\mu}(\alpha)r_j(\alpha)
\left \< {\bf \Lambda_i}
|\Gamma_{\mu}|
{\bf \Lambda_j}
\right \>
\\ \nn
&-&
\sum_{\alpha,\beta}
\sum_{\mu,\nu}
\sum_{\br_k}\sum_{c,c'=\{a,s\}}
r_i(\alpha)\bar a_{\mu}(\alpha)
u_k^c(\alpha,\beta)
v_k^{c'}(\alpha,\beta)\bar a_{\nu}(\beta)r_j(\beta)
\left \< {\bf \Lambda_i}
|\Gamma_{\mu}\Gamma_{cc'}\Gamma_{\nu}|
{\bf \Lambda_j}
\right \>+...\\
&=&C_{ij}^{(0)}+C_{ij}^{(1)}+...
\ee
The structure of this expansion make possible a further diagrammatic expansion 
of the quantities $C_{ij}$ as a series of term $C_{ij}^{(k)}$, with $k\ge 0$. 
The first term $C_{ij}^{(0)}$ has symmetry factors
\bb\nn
\< {\bf \Lambda_i}|\Gamma_{x}|{\bf \Lambda_j}\>
&=&
\overbrace{i^{m_i+n_i+m_j+n_j}}^{\text{$c_{ij}$}}\overbrace{\left [ (-1)^{m_i}-(-1)^{m_j} \right ]
\left [(-1)^{n_i}+(-1)^{n_j} \right ]}^{\text{$\gamma^{(1)}_{ij}$}}:=c_{ij}\gamma^{(1)}_{ij},
\\ \label{Sij1}
\< {\bf \Lambda_i}|\Gamma_{y}|{\bf \Lambda_j}\>
&=&
\underbrace{i^{m_i+n_i+m_j+n_j}}_{\text{$c_{ij}$}}\underbrace{\left [ 1+(-1)^{m_i+m_j}\right ]
\left [(-1)^{n_j}-(-1)^{n_i} \right ]}_{\text{$\gamma^{(2)}_{ij}$}}:=c_{ij}\gamma^{(2)}_{ij}.
\ee
It is easy to see that $\< {\bf\Lambda_i}|\Gamma_{x}|{\bf \Lambda_j}\>=0$ for 
monomers on the boundary or on the same column, when $m_i=m_j$. {\it A contrario}, for pairs of monomers on the same line 
$n_i=n_j$, we have $\< {\bf \Lambda_i}|\Gamma_{y}|{\bf \Lambda_j}\>=0$.
The first term $C_{ij}^{(0)}$ can be expressed in the discrete case as 
\bb\label{exact-bord}
C_{ij}^{(0)}(L)&=&\frac{2c_{ij}}{(L+1)^2}\sum_{p,q=1}^{L/2}
\Big \{
\frac{\gamma^{(1)}_{ij} t_x\cos\frac{\pi p}{L+1}}{t_x^2\cos^2\frac{\pi 
p}{L+1}+t_y^2\cos^2\frac{\pi q}{L+1}}+
\frac{\gamma^{(2)}_{ij} t_y\cos\frac{\pi q}{L+1}}{t_x^2\cos^2\frac{\pi 
p}{L+1}+t_y^2\cos^2\frac{\pi q}{L+1}}\Big \}
\nn\\
&\times&
\sin\frac{\pi p m_i}{L+1}\sin\frac{\pi p m_j}{L+1}
\sin\frac{\pi q n_i}{L+1}\sin\frac{\pi q n_j}{L+1}.
\ee
This expression is valid for 2 monomers on any of the four boundaries of the lattice, 
and is identical, when $m_i=m_j=L$, to expression obtained for the same ($m_{i}=n_{i}=L$) boundary case \eref{CijBound}. Indeed the first order of the expansion \eref{Wexp} is valid only on the boundaries where the matrix $W$ is actually equal to the matrix $M$.
One could demonstrate that this 2-point correlation is actually exact $C_{ij}^{(0)}=C_{ij}$ for boundary monomers because of the cancelation of higher terms in the perturbative expansion, accordingly the expression \eref{exact-bord} is a general exact result for any positions anywhere on the four boundaries. Therefore, it will be very efficient to use this exact closed-form to evaluate scaling behaviors of correlation functions between monomers on the surface and at the corners. This present perturbative expansion can be performed to the next leading order to evaluate bulk correlations, and will be detailed elsewhere.

\subsection{Scaling behavior of monomer correlation functions}

\begin{figure}[!h]
\includegraphics[scale=1.15]{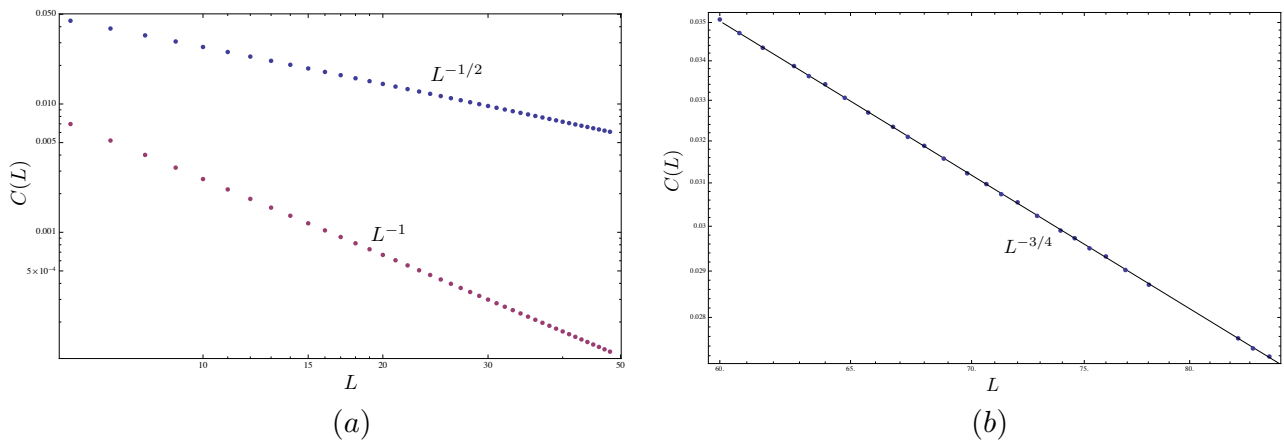}
\caption{($a$) Bulk-bulk and surface-surface correlation functions. ($b$) Bulk-surface monomer correlation functions}
\label{correlation-mon}
\end{figure}

\begin{figure}[!h]
\center{
\includegraphics[scale=0.9]{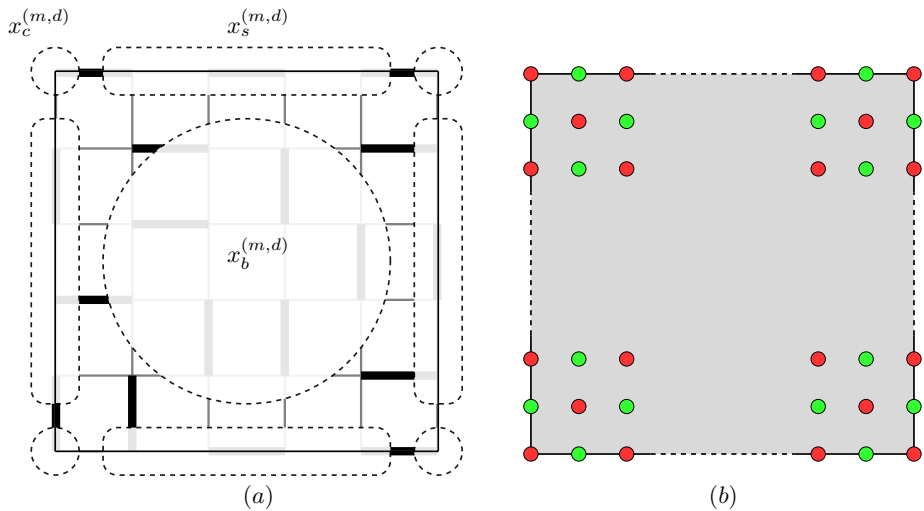}
}
\caption{($a$) Bulk, surface and corner decomposition of the square. ($b$) Representation of the values of the scaling dimensions of monomer operators close to the corners. Monomers on red sites have dimension $1/2$ while monomers on green sites have dimension $3/2$. Let us notice that this distinction is different from the even/odd sublattice distinction because opposite sublattice sites may have the same dimension and {\it vice verca}.}
\label{cornercorr2}
\end{figure}

\begin{figure}[!h]
\center{
\includegraphics[scale=1.1]{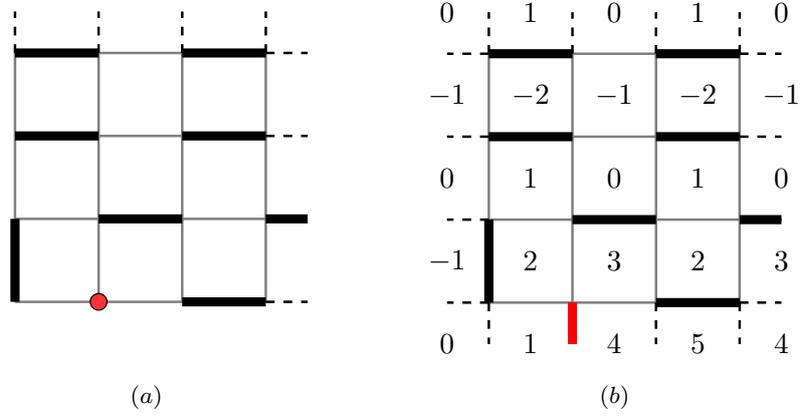}
\caption{($a$) Representation of the corner region of a lattice with one monomer at the corner. ($b$) The monomer can be represented by a virtual red dimer which gives the same configurations. The different values of the height field either sides of the corner are induced by a {\it bcc} operator of scaling dimension $25/32$.}
\label{heightmap3}
}
\end{figure}

\begin{figure}[!h]
\center{
\includegraphics[scale=0.7]{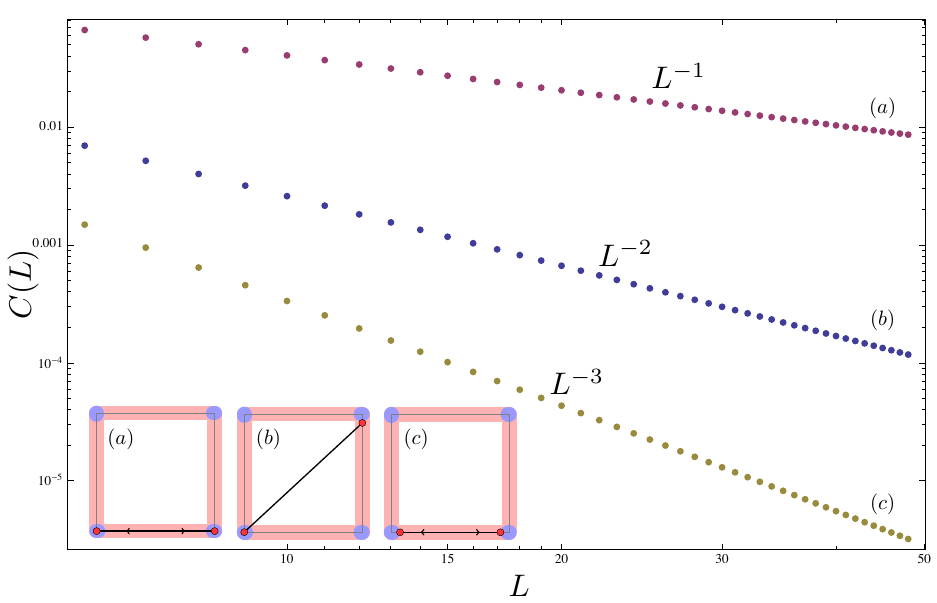}}
\caption{Corner-Corner in the three different situations pictured on the graphical representation: ($a$) the two monomers exactly on the corners. ($b$) One monomer on one corner and the other one on a adjacent site of another corner. ($c$) Two monomers on a adjacent sites of opposite corners.}
\label{cornercorr}
\end{figure}

\begin{figure}[!h]
\center{
\includegraphics[scale=0.9]{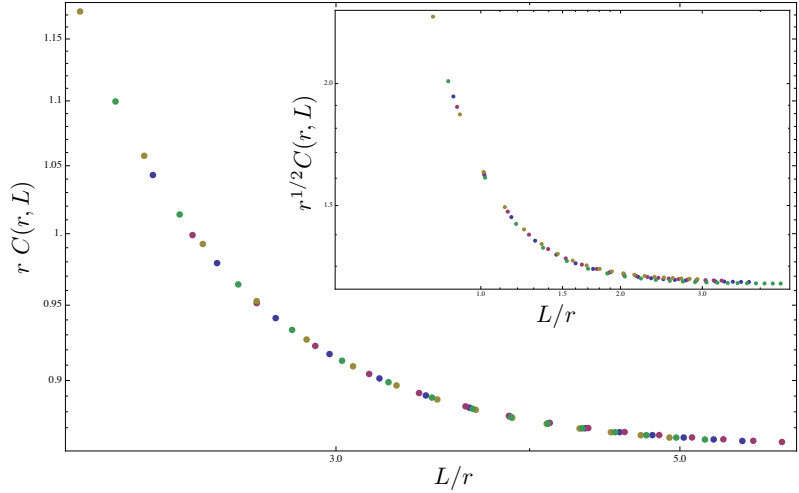}
}
\caption{Finite size scaling of monomer surface-surface and bulk-bulk (inset) correlation functions in log-log scale for $L=100,200,300,400$. The surface correlations are measured far from the corners and the bulk correlations are measured far from boundaries and corners.}
\label{scaling-dimer}
\end{figure}

Here, we shall analyze monomer-monomer correlation functions using our pfaffian solution detailed previously and compared to the Coulomb gas interpretation of the dimer model. As we saw in section \ref{sect1}, the dimer model on a rectangular geometry admit a {\it bcc} operator on every of the four corners, and it has to be taken into account for the analysis of the scaling dimension operators, in particular for corner correlations. Indeed in the case of monomers deep in the bulk\footnote{far from surfaces and corners.} or deep in the surface\footnote{far from corners.}, the scaling dimensions are respectively $x^{(m)}_{b}=1/4$ and $x^{(m)}_{s}=1/2$, leading to the following scaling of correlation functions (see \efig{cornercorr2}($a$))
\bb\label{dimerbulk}
{\rm\ monomer \ correlations}\rightarrow \left\{
    \begin{array}{ll}
       $bulk-bulk behavior$ \rightarrow C(L)\sim L^{-2x^{(m)}_{b}}\sim L^{-1/2}\\
$surface-surface behavior$ \rightarrow C(L)\sim L^{-2x^{(m)}_{s}}\sim L^{-1}\\
$bulk-surface behavior$ \rightarrow C(L)\sim L^{-x^{(m)}_{s}-x^{(m)}_{b}}\sim L^{-3/4}.
    \end{array}
\right.
\ee
These known results are in perfect agreement with our exact solution ({\it cf.} \efig{correlation-mon}) where we fixed the positions of two monomers for increasing system size $L$ (all the correlations are measured for $t_{x}=t_{y}=1$).  
The behaviors of bulk and surface monomer correlation functions had already been studied in several papers, and the scaling dimensions are related to the scaling dimensions of operators of the $2d$ Ising model {\it via} the expression of monomer-monomer correlations as spin-spin correlations \cite{perk84}. At present, we consider the effects of corners in our system, which seems to be more difficult to consider as we have seen for the corner contribution to the free energy. Fortunately conformal invariance predicts a relation for the scaling dimension of an operator in the vicinity of a corner of an angle $\theta$ in terms of the scaling dimension of the same operator on the surface \cite{cardy1988finite}
\bb\label{cftcorner}
x_{c}=\frac{\pi}{\theta}x_{s}.
\ee
If we believe in this formula, we should obtain the value $x^{(m)}_{c}=1$ for $\theta=\pi/2$, which leads to the behavior $C(L)\sim L^{-2}$ for corner-corner correlation functions. This result contradicts our exact evaluation, where the exponent seems to change according to the exact location of the monomers (see \efig{cornercorr2}($b$) and \efig{cornercorr}), and where three different cases arise. Unlike the surface and bulk cases where the scaling dimensions are uniquely defined, the corner scaling dimension appears to be less trivial to analyze, and the influence of the {\it bcc} operators has to be carefully taking into account. We should mention that the same kind of analysis has been done for the Ising model, where the magnetization was measured for various spins close to a corner \cite{peschel1985some}.
We saw previously \eref{coin2}, that {\it bcc} operators add a logarithm term in the expression of the free energy $\mathcal{F}$ of a rectangular system, then this contribution to the partition function scale as
\bb
Q(L)\sim L^{-2\big(h^{(1)}_{{\it bcc}}+h^{(2)}_{{\it bcc}}+h^{(3)}_{{\it bcc}}+h^{(4)}_{{\it bcc}}\big)}= L^{-2(1/32+1/32+1/32+1/32)}.
\ee
Indeed putting two monomers exactly on the corner of the same boundary (see \efig{cornercorr}($a$)) is equivalent to a height shift of value $3\pi/2$ in each corner. This height shift is induced by an operator of dimension $9/32$, leading to the following behavior of the partition function with the two monomers
\bb
Q_{2}(L)\sim L^{-2\big(1/32+1/32+9/32+9/32\big)}.
\ee
Here the correlation function scale then as $C(L)=Q_{2}Q_{0}^{-1}\sim L^{-1}$ leading to the value $x^{(m)}_{c}=1/2$ of the monomer corner scaling dimension. Nevertheless, if one choose the diagonal corners (see \efig{cornercorr}($b$)), we place a monomer on the first corner which is again equivalent to the insertion of a {\it bcc} operator of dimension $9/32$ (see \efig{heightmap2}) and the other one on a neighboring site of the other corner which is equivalent to the insertion of a {\it bcc} operator of dimension $25/32$ (see \efig{heightmap3}), we found the behavior $C(L)\sim L^{-2}$. 
Finally, if both of the two monomers are on a neighboring site of a corner (see \efig{cornercorr}($c$)), then the correlation function is $C(L)\sim L^{-3}$. This three different situations are summarized in \efig{cornercorr}, showing that our exact computations are in perfect agreements with Coulomb gas predictions for the behavior of correlation functions in the vicinity of a corner. A more general statement is that the monomer scaling dimension near a corner depends crucially of the sublattice considered as explained in \efig{cornercorr2}($b$). This phenomena leads to two different values of the scaling dimension for corner monomers $x^{(m)}_{c}=1/2\ \mathrm{or}\ 3/2$ which is in agreement with the CFT formula \eref{cftcorner} in average when lattice effects are forgotten. 
A general study of the finite size behavior of correlation functions can be performed as well, leading to the following scaling ansatz 
\bb\label{scaling1}
C(r_{1},r_{2},L)=\abs{r_{1}-r_{2}}^{-x_{1}-x_{2}}\Phi(\abs{r_{1}-r_{2}}^{-1}L),
\ee
where $r_{1}$ and $r_{2}$ are the positions of the two monomers, with respective scaling dimensions $x_{1}$ and $x_{2}$. The scaling function $\Phi(u)$ depends on the position of the operators and goes to a constant in the scaling limit $u\rightarrow\infty$ (see \efig{scaling-dimer}). The translation and rotational invariance has been checked analytically and numerically, and in the following we will use $\abs{r_{1}-r_{2}}=r$
\bb
C(r)\sim r^{-x_{1}-x_{2}}.
\ee
This scaling behavior is shown for bulk-bulk and surface-surface correlations in \efig{scaling-dimer}.
The exact form of the scaling function $\Psi$ seems hard to obtain explicitly, but at least for boundary and corner cases it should be possible to extract the scaling behaviors using the expression \eref{exact-bord} of exact correlation functions. We let this question for a future work and we hope that comparisons with CFT predictions can be made.

\subsection{Scaling behavior of dimer correlation functions}

As we have shown in section \ref{sect1} of this article, the bulk correlation between a dimer covering the two neighboring sites $i$ and $j$ and another dimer covering the two neighboring sites $m$ and $n$ can be computed in the Kasteleyn-Fisher-Temperley pfaffian formalism leading to a behavior in $L^{-2}$ in the thermodynamic limit. Let us note $D(L)$ this quantity. In the Coulomb gas approach, dimers are interpreted as electric charges with scaling dimensions $x^{(d)}_{b}:=x_{\frac{1}{4\pi}}(1,0)$, the result for dimer-dimer correlations gives us the value $x^{(d)}_{b}=1$. Actually it is straightforward to study dimer correlations here. Indeed a dimer can be seen as two neighboring monomers, thus a dimer-dimer correlation is simply a 4-point monomer-monomer correlation, which can be evaluated with our solution. In the following, one shows how to construct the dimer-dimer correlation in the boundary case, for the bulk case the situation is essentially the same but expressions are less convenient. Explicitly the correlation between two dimers at position $(r_{i},r_{j})$ and $(r_{m},r_{n})$ is 
\bb
Q_{4}(r_{i},r_{j},r_{m},r_{n})Q^{-1}_{0}=C_{ij}C_{mn}-C_{im}C_{jn}+C_{in}C_{jm}.
\ee
If we choose that $r_{i}$ and $r_{m}$ are on the same sublattice (then $r_{j}$ and $r_{n}$ are on the other one), a straightforward consequence is that the second term $C_{im}C_{jn}$ vanishes, moreover the first term $C_{ij}C_{mn}$ tends to a constant in the thermodynamic limit in such way that we can define the dimer-dimer correlation function as
\bb\label{dimcor}
Q_{4}(r_{i},r_{j},r_{m},r_{n})Q^{-1}_{0}-C_{ij}C_{mn}=C_{in}C_{jm}\sim D(L).
\ee
In this way, all the configurations of dimer correlations are available, and all the scaling dimensions of electric charges (bulk, surface, corner) may be analyzed {\it stricto sensu} and compared with the Coulomb gas theory.  We can show that the well known bulk behavior is recovered very precisely, furthermore, surface and corner correlations may be examined as well leading to the following behaviors 
\bb\label{dimerbulk}
{\rm\ dimer \ correlations}\rightarrow \left\{
    \begin{array}{ll}
 $bulk-bulk behavior$ \ \rightarrow D(L)\sim L^{-2x^{(d)}_{b}}\sim L^{-2},\\
$surface-surface behavior$ \rightarrow D(L)\sim L^{-2x^{(d)}_{s}}\sim L^{-2},\\
$corner-corner behavior$\rightarrow D(L)\sim L^{-2x^{(d)}_{c}}\sim L^{-4}.
    \end{array}
\right.
\ee
Unlike monomer correlations, dimer correlations are much easier to interpret in the Coulomb gas framework. Indeed, the absence of additional change of boundary conditions\footnote{Of course the four corner {\it bcc}'s with dimension $1/32$ are still present but do not play any role in dimer-dimer correlation functions.} in the partition function allows for a direct determination of dimer scaling dimensions. The particular form of dimer correlations \eref{dimcor} predicts that  $x^{(d)}_{s}=2x^{(m)}_{s}=1$\footnote{Let us notice that the fact that $x^{(d)}_{b}=x^{(d)}_{s}$ is a pure coincidence.}  and $x^{(d)}_{c}=x^{(m)}_{c_{+}}+x^{(m)}_{c_{-}}=2$\footnote{A dimer in the corner is formed by two neighboring monomers with dimension $x^{(m)}_{c_{+}}$ and $x^{(m)}_{c_{-}}$.}. We notice here that the formula \eref{cftcorner} checked out in that case. A careful and detailed study of surface and corner operators has to be performed to unravel this point. The scaling form of correlation functions \eref{scaling1} holds in the dimer case as well, using the dimer scaling dimensions (see Table \ref{scaling-dimension}). 
\begin{table}[h!]\center
\begin{tabular}{|c|cccc|} \hline
 scaling dimension ($g_{\rm{free}}=1/4\pi$) & ~bulk~ & ~surface~   & ~corner~ & \\ \hline
$x^{(d)}$   & $1$ & $1$  & $2$ &     \\ 
$x^{(m)}$   & $1/4$   & $\textcolor{black}{1/2}$  & $1/2$ or $3/2$ &        \\ \hline
\end{tabular}
\caption{Bulk, surface and corner values of dimer and monomer scaling dimensions for the free ($g_{\rm{free}}=1/4\pi$) fixed point. The corner monomer scaling dimension depends of its exact location.}
\label{scaling-dimension}
\end{table}
The solution presented in this article can be also used to calculate more complex correlation functions, combining dimer and monomer scaling dimensions. {\it A posteriori}, more complicated object like trimers, quadrimers or more generally, string of $k$ neighboring monomers ($k$-mer) can be studied as well, which correspond to various charged particles in the Coulomb gas formalism.

\section{About some combinatorial properties}\label{sect4}

In this section, one shows a curious combinatorial analogy between the partition function of the close packing dimer model on a $L\times L$ square lattice with open boundary conditions, and the same partition function  with boundary monomers. One start reminding some properties of the pure dimer model partition function, and we show, thanks to our exact calculation of the partition function with $2n$ monomers, that this analogy can be understood and demonstrated. Hereinafter, the Boltzmann weights $t_{x}$ and $t_{y}$ are taken to be the unity in such way that the partition function is exactly equal to the perfect matching number. All the results presented in this section has been checked with depth-first \cite{krauth2006statistical} algorithms up to size $L=10$. For bigger sizes, Monte-carlo simulations \cite{krauth2003pocket} or transfer matrix calculation \cite{lieb1967} has to be implemented. 
\subsection{Partition function without monomers}
\begin{figure}[h!]
\center{
\includegraphics[scale=1.2]{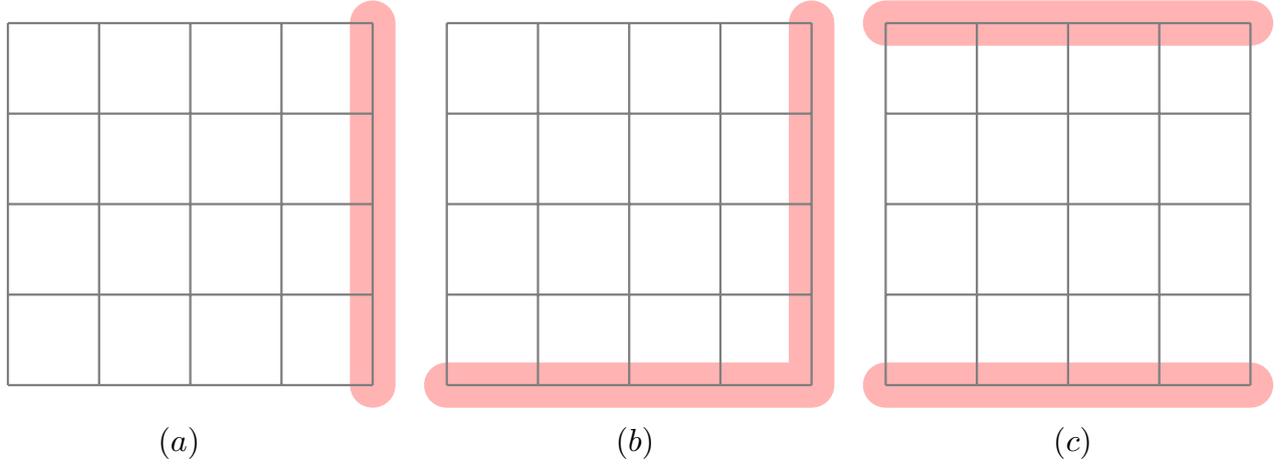}
}\caption{The three different possibilities of two monomer's location. ($a$) on a same boundary ($b$) on adjacent boundaries or ($c$) on opposite boundaries.}
\label{geometry}
\end{figure}

The partition function of the pure dimer model on a $M\times N$ lattice with open boundary conditions is 
\bb
{Q}_0(M,N)=\prod_{p=1}^{M/2}\prod_{q=1}^{N/2}\left [ 4\cos^2\frac{\pi 
p}{M+1}+4\cos^2\frac{\pi q}{N+1} \right ],
\ee
which can be written for the special case of the square geometry $M=N=L$
\bb\label{comb}
\boxed{{Q}_0(L)=2^{L/2}.g^{2}_{L/2}}
\ee
where $g_{L/2}$ is a 
number sequence (OEIS A065072) \footnote{The On-Line Encyclopedia of Integer Sequences \url{https://oeis.org/}} equal, for $L=2,4,6,8,10,12,14...$, to
\bb\nn
g_{L/2}=\{1, 3, 29, 901, 89893, 28793575, 29607089625...\}.
\ee
The resulting sequence for the partition function is then (OEIS A004003) for $L=2,4,6,8,10,12,14$
\bb\nn
Q_{0}=\{2, 36, 6728, 12988816, 258584046368, 53060477521960000,112202208776036178000000...\}.
\ee
For example, the number of configurations of dimers on the chessboard ($L=8$) is $Q_{0}(8)=2^{4}g^{2}_{4}=2^{4}\times \textcolor{black}{901}^{2}=12988816$ as previously noticed by Fisher \cite{fisher1961statistical}. Another observation is that the number of configuration on the square $L\times L$ is always even. It is less trivial to notice that $\{g_{p}\}$ is a sequence of odd number satisfying the relation \cite{john2000strange}
\bea
g_{p}&=&p+1({\rm mod} \ 32)  \ \ {\rm if} \ \ p \ \ {\rm even} \nonumber\\
&=&(-1)^{(p-1)/2}\times p ({\rm mod} \ 32) \ \ {\rm if} \ \ p \ \  {\rm odd}.
\eea
The exact solution of the dimer model with one boundary monomer allows for the same kind of number theory analysis ({\it cf.} \cite{kong2006packing} for details). The aim of the following sections is to look in more details at the form of the partition function of a dimer model of on a $L\times L$ square ($L$ even) lattice with $2n$ monomers. One allows the $2n$ monomers to be anywhere on the four boundaries of the square (see \efig{geometry}).

\subsection{Partition function with two boundary monomers}
We saw previously that the expression of this partition function $Q_{2n}$ is related to the pure dimer model $Q_{0}$ by the formula
\bb\label{eqbord}
Q_{2n}=Q_{0}.\Pf(C),
\ee
where the size of the matrix $C$ depend of the number of monomers. Previously, the partition $Q_{0}$ has been shown
to possess a remarkable expression \eref{comb} and we would like to determine whether or not, the partition function $Q_{2n}$ 
admit the same kind of properties.
In the case of two monomers anywhere on the boundaries, we saw that $W=M$ and then the expression \eref{eqbord} reduces to
\bb\label{eqbord1}
Q_{2}=Q_{0}.C_{ij}.
\ee
\subsubsection{Inline boundary monomers}
Initially, we choose to restrict the monomers to live on the same boundary $m_{i}=m_{j}=L$ with $n_{i},n_{j}\in [1,L]$  ({\it cf.} \efig{geometry}(a)). In that particular situation the matrix elements $C_{ij}$ take the following form \eref{CijBound}
\bb
C_{ij}=R_{ij}\sqrt{\frac{2}{L+1}}\sum_{p,q=1}^{L/2}
\frac{i^{1+n_i+n_j}\cos\frac{\pi q}{L+1}\sin^2\frac{\pi p}{L+1}}
{\cos^2\frac{\pi p}{L+1}+\cos^2\frac{\pi q}{L+1}}
\sin\frac{\pi q n_i}{L+1}\sin\frac{\pi q n_j}{L+1}
\ee
where
\bb
R_{ij}&=&\pm 2 \ \ \mathrm{if} \ n_{i}\in\mathbb{Z}_{2p}(\mathbb{Z}_{2p+1}) \ \mathrm{and}  \ n_{j}\in
\mathbb{Z}_{2p}(\mathbb{Z}_{2p+1})\nonumber\\
&=&0 \ \ \mathrm{if} \ \ n_{i}\in\mathbb{Z}_{2p} \ \mathrm{and} \ n_{j}\in\mathbb{Z}_{2p+1} \ \mathrm{or \ converserly}.
\ee
In table \ref{table2}, we evaluate this expression using $\textsc{Mathematica}^{\circledR}$, restricting one monomer to be in $n_{j}=1$ and the second to be between $1$ to $L$ for several system sizes. 
\begin{table}[h!]\center
\begin{tabular}{|c|cccccc|} \hline
$C_{ij}(L)$  & ~$4$~   & ~$6$~  & ~$8$~ & ~$10$~ & ~$12$~& ~$14$~ \\ \hline
$(1,1),(2,1)$   & $1/2 $  & $1/2$ & $1/2$                     & $1/2$     & $1/2$ & $1/2$       \\
$(1,1),(4,1)$     & $1/3$  & $9/29$ & $275/901$                     & $27293/89893$     & $8724245/28793575$  & $8962349805/29607089625$      \\
$(1,1),(6,1)$     & $.$  & $7/29$ & $199/901$                     & $19279/89893$     & $6103405/28793575$  & $6242309595/29607089625$     \\ 
$(1,1),(8,1)$     & $.$  & $.$ & $169/901$                     & $15395/89893$     & $4750015/28793575$     & $4800013155/29607089625$  \\ 
$(1,1),(10,1)$     & $.$  & $.$ & $.$                     & $13761/89893$     & $4036195/28793575$    & $3979640565/29607089625$   \\ 
$(1,1),(12,1)$     & $.$  & $.$ & $.$                     & $.$     & $3721985/28793575$   & $3520442385/29607089625$    \\
$(1,1),(14,1)$    & $.$  & $.$ & $.$                     & $.$     & $.$    & $3311911215/29607089625$   \\ \hline
$g_{L/2}$     & $3$  & $29$ & $901$                     & $89893$     & $28793575$    & $29607089625$   \\ \hline
\end{tabular}
\caption{Correlation function $C_{ij}$ for a boundary monomer 
($m_i=m_j=L$) fixed on the first site $n_j=1$ as function of the 
ordinate $n_{i}$ for several system sizes $L$ (see \efig{geometry}($a$)). The value of $C_{ij}$ where the two monomers are on the corner $(1,1),(2,1)$ is always equal to $1/2$, because it is equivalent to force a dimer to be on the corner and then split the number of configuration by two. Bottom line: Values of the sequence $g_{L}$ for $L=4..14$.}
\label{table2}
\end{table}
One can observe that there is a curious relation between the expression $C_{ij}$ and the sequence $g_{L/2}$ present in the partition function $Q_{0}$, more precisely one can deduce a proportionality relation 
\beq
C_{ij}(L)\propto g^{-1}_{L/2},
\eeq
which appears to be valid for all system sizes $L$ in the case of inline monomers.
\subsubsection{General case}
The general expression of the matrix elements of the correlations between boundary monomers \eref{exact-bord}, valid in all the geometries of \efig{geometry} can be written as
\bb\nn
C_{ij}&=&\frac{2c_{ij}}{(L+1)^2}\sum_{p,q=1}^{L/2}
\Big \{
\frac{\gamma^{(1)}_{ij} t_x\cos\frac{\pi p}{L+1}}{t_x^2\cos^2\frac{\pi 
p}{L+1}+t_y^2\cos^2\frac{\pi q}{L+1}}+
\frac{\gamma^{(2)}_{ij} t_y\cos\frac{\pi q}{L+1}}{t_x^2\cos^2\frac{\pi 
p}{L+1}+t_y^2\cos^2\frac{\pi q}{L+1}}\Big \}
\\
&\times&
\sin\frac{\pi p m_i}{L+1}\sin\frac{\pi p m_j}{L+1}
\sin\frac{\pi q n_i}{L+1}\sin\frac{\pi q n_j}{L+1}.
\ee
In table \ref{table3} , we evaluate this expression for the two other geometries.  
\begin{table}[h!]\center
\begin{tabular}{|c|ccccc|} \hline
$C_{ij}(L)$ & ~$4$~   & ~$6$~  & ~$8$~ & ~$10$~ & ~$12$~ \\ \hline
$(1,1),(L,1)$  & $1/3$  & $7/29$ & $169/901$                     & $13761/89893$     & $3721985/28793575$    \\
$(1,1),(L,3)$   & $1/(2\times3)$   & $5/29$  & $138/901$ & $12127/89893$        & $3407775/28793575$   \\
$(1,1),(L,5)$     & $.$  & $5/(2\times29)$ & $95/901$                     & $9475/89893$     & $2864755/28793575$     \\ 
$(1,1),(L,7)$     & $.$  & $.$ & $95/(2\times901)$                     & $6389/89893$     & $2194565/28793575$    \\ 
$(1,1),(L,9)$     & $.$  & $.$ & $.$                     & $6389/(2\times89893)$     & $1471805/28793575$      \\ 
$(1,1),(L,11)$   & $.$  & $.$ & $.$                     & $.$     & $1471805/(2\times28793575)$       \\ \hline
$C_{ij}(L)$ & ~$4$~   & ~$6$~  & ~$8$~ & ~$10$~ & ~$12$~ \\ \hline
$(1,1),(L,1)$  & $1/3$  & $7/29$ & $169/901$                     & $13761/89893$     & $3721985/28793575$    \\
$(1,2),(L,2)$   & $1/6$   & $2/29$  & $30/901$ & $1634/89893$        & $314210/28793575$   \\
$(1,3),(L,3)$     & $1/6$  & $9/29$ & $125/901$                     & $11109/89893$     & $3178965/28793575$     \\ 
$(1,4),(L,4)$     & $1/3$  & $9/29$ & $155/901$                     & $4720/89893$     & $984400/28793575$    \\ \hline
$g_{L/2}$     & $3$  & $29$ & $901$                     & $89893$     & $28793575$      \\ \hline
\end{tabular}
\caption{Top: Correlation function $C_{ij}$ between a monomer at position
($m_i=n_i=1$) and another at position ($m_j=L, n_j$) for several system sizes $L$ and for $n_{j}=1,3,5,7,9,11$ ({\it cf.} \efig{geometry}($b$)). We notice that the expression of the last line in each row is the half of the expression of the penultimate line. Bottom: Correlation function $C_{ij}$ between opposite side monomers for several system sizes $L$ and for $n_{i}=n_{j}=1,2,3,4$ (see \efig{geometry}($c$)).}
\label{table3}
\end{table}
The same relation holds in this case as well, and we conjecture that the expression of the 2-point correlation takes the following form
\beq\label{cij}
C_{ij}(L)=\alpha^{(2)}_{ij}(L)g^{-1}_{L/2},
\eeq
no matter the positions of the two monomers on the boundaries, where $\alpha^{(2)}_{ij}$ depends only of the positions of the two monomers and of the system size $L$. Consequently, the partition function of the dimer model with two boundary monomers reads
\bb
\boxed{
Q_{2}(L)=\alpha^{(2)}_{ij}(L).g_{L/2}
}\ee 

\subsection{Partition function with $2n$ boundary monomers}

It is worth looking at higher number of monomers to conjecture a more general form of the partition function. We have conjectured that the matrix elements of the correlation matrix are proportional to the sequence $g_{L/2}$, thus thanks to the general pfaffian solution with $2n$ monomers \eref{eqbord}, we obtain the formulas
\bb\nn
Q_{2}&=&Q_{0}.C_{ij},\nn\\
Q_{4}&=&Q_{0}.\big(C_{ij}C_{kl}-C_{ik}C_{jl}+C_{il}C_{jk}\big),\nn\\
Q_{6}&=&Q_{0}.\big(C_{ij}C_{kl}C_{mn}-C_{il}C_{jl}C_{mn}+C_{il}C_{jm}C_{kn}-...\big),\\
&\vdots&\nonumber
\ee
where the pure partition function takes the form \eref{comb}, therefore the partition functions are proportional to power of $g_{L/2}$
\bb
Q_{0}(L)&=&2^{L/2}.g^{2}_{L/2},\nonumber\\
Q_{2}(L)&=&\alpha^{(2)}_{ij}(L).g_{L/2},\nonumber\\
Q_{4}(L)&=&\alpha^{(4)}_{ijkl}(L).g^{0}_{L/2},\nonumber\\
Q_{6}(L)&=&\alpha^{(6)}_{ijklmn}(L).g^{-1}_{L/2},\\
&\vdots&\nonumber
\ee
which can be generalized for $2n$ monomers at positions $i_{1}, i_{2},..., i_{2n}$
\bb
Q_{2n}(L)=\alpha^{(2n)}_{i_{1}i_{2}...i_{2n}}(L).g^{2-n}_{L/2},
\ee
in such way that a relation between $Q_{2p}$ et $Q_{2q}$ can be founded ($ p,q \geqslant 1$), dropping all the indices but $p$ and $q$ for simplicity, we found  {\it ex hypothesi}
\beq
\boxed{
\frac{Q_{2p}}{Q_{2q}}=\frac{\alpha^{(2p)}}{\alpha^{(2q)}}g^{q-p},}
\eeq
valid for $2p$ and $2q$ monomers anywhere on the boundaries of the square lattice. Finally all these numerical relations between dimer partition functions with and without boundary monomers are the consequence of \eref{eqbord} and \eref{cij}, which are unfortunately no longer valid for bulk monomers.

\section{Conclusion}

In this work the classical dimer model was discussed in great details both in a fermionic and bosonic field theory formulation. The bosonic formulation of the dimer model is based on the so-called height mapping and it is well suited for phenomenological predictions about correlations between dimers and monomers in a Coulomb gas context.
Then we presented a practical and complete fermionic solution of the $2d$ dimer
model on the square lattice with an arbitrary number of monomers. Furthermore, the Tzeng-Wu solution of the dimer model with a boundary monomer was found to be included in our theory. Interpretations of finite size effects of the Tzeng-Wu solution in a CFT/Coulomb gas framework has been performed, and we showed that a careful examination of boundary conditions in the model allowed us to recover the central charge of the free fermion/free boson field theory.
The exact expression of correlation functions between monomers has been written in terms of the product of two pfaffians, and we gave an explicit formula for boundary correlations valid for the four boundaries of the rectangle. This solution has been used to compute correlations for several configurations in order to extract bulk, surface and corner scaling dimensions for dimer and monomer operators. All these results were interpreted in the Coulomb gas formalism, and we showed that all the predictions of the CFT were in accordance with a $c=1$ theory.
Last but not least, the exact closed-form expression of correlations between boundary monomers has been extensively used to extract some combinatorial and numerical informations about the partition function of the model. Furthermore, an unexpected relation has been found between partition functions with and without boundary monomers, and has been demonstrated thanks to our pfaffian solution.
Generally, this Grassmann method can also be used for studying more general correlation functions, thermodynamical 
quantities, or transport phenomena of monomers. Other types of lattices, such as hexagonal lattice and other boundary conditions, can also be considered, as well as more precise comparisons with CFT results about rectangle geometry \cite{stephan2014emptiness}. The same analysis of corner contribution to free energy as well as critical exponents can be studied in the interacting dimer model using the height mapping and results will be presented elsewhere.
A future challenge emerging out of this present work is the study of other two dimensional dimer related models as the trimer model \cite{ghosh2007random} or the four-color model \cite{kondev1995four,fjaerestad2012dimer} which can be seen as an interacting colored dimer model. Work in those directions is in progress.

\newpage

\begin{appendices}

\section{Grassmann variables, determinant permanent and all that}\label{grassmann}

A $n$-dimensional Grassmann algebra \footnote{The presentation closely follows \cite{itzykson1991statistical}.} is the algebra generated by a set of variables $\{a_{i}\}$, with $i=1..n$ satisfying
\bb
\{a_{i},a_{j}\}=0,
\ee
{\it i.e.} they anti-commute, which implies in particular that $a_{i}^{2}=0$. The algebra generated by these quantities contains all expressions of the form
\bb
f(a)&=&f^{(0)}+\sum_{i}f^{i}a_{i}+\sum_{i<j}f^{ij}a_{i}a_{j}+..\nonumber\\
&=&\sum_{0\leqslant p \leqslant n}\sum_{i}\frac{1}{p!}f^{i_{1}...i_{p}}a_{i_{1}}a_{i_{2}}...a_{i_{p}},
\ee
where the coefficients are antisymmetric tensors with $p$ indices, each ranging from $1$ to $n$. Since there are  $\binom{n}{p}$ such linearly independent tensors, summing over $p$ from $0$ to
$n$ produces a $2n$-dimensional algebra. The anticommunting rule allows us to define an
associative product
\bb
f_{1}(a)f_{2}(a)=f_{1}^{0}f_{1}^{0}+\sum_{i}(f_{1}^{0}f_{2}^{i}+f_{1}^{i}f_{2}^{0})a_{i}+\frac{1}{2}\sum_{ij}(f_{1}^{ij}f_{2}^{0}+f_{1}^{i}f_{2}^{j}-f_{1}^{j}f_{2}^{i}+f_{1}^{0}f_{2}^{ij})a_{i}a_{j}+..
\ee
Please note that in general $fg$ is not equal to $\pm gf$. Nevertheless the subalgebra containing terms with an even number (possibly zero) of $a$ variables commutes with any element $f$. Having defined sum and products in the Grassmann algebra we now define a left derivative $\partial_{i}:=\partial_{a_{i}}$. The derivative gives zero on a monomial which does not contain the variable $a_{i}$. If the monomial does contain $a_{i}$, it is moved to the left (with the appropriate sign due to the exchanges) and then suppressed. The operation is extended by linearity to any element of the algebra. A right derivative can be defined similarly. From this definition the following rules can be obtained
\bb
&&\{\partial_{i},\partial_{j}\}=0\nonumber\\
&&\{\partial_{i},a_{j} \}=\delta_{ij}.
\ee
Integrals are defined as linear operations over the functions $f$ with the property that they can be identified with the (left) derivatives \cite{berezin}. Correspondingly
\bb
&&\int \dd a_{i} \ f(a)=\partial_{i}f(a),\nonumber\\
&&\int \dd a_{i}\dd a_{j} \ f(a)=\partial_{i}\partial_{j}f(a),
\ee
which leads to the generalization
\bb
\int \dd a_{i_{k}}\dd a_{i_{k-1}}...\dd a_{i_{1}} \ f(a)=\partial_{i_{k}}\partial_{i_{k-1}}...\partial_{i_{1}}f.
\ee
It is obvious that this definition fulfills the constraint of translational invariance
\bb
\int \dd c (c_{1}+c_{2}a)=\int \dd c  [c_{1}+c_{2}(a+b)],
\ee
which requires
\bb
\int \dd a_{i} \ a_{j}=\delta_{ij}.
\ee
Changes of coordinates are required to preserve the anti-commuting structure of the Grassmann algebra, this allows non-singular linear transformations of the form $b_{i}=\sum_{j}A_{ij}a_{i}$. One then can verify that by setting $f(a)=F(b)$ one can obtain the following relation
\bb
\int\prod_{i} \dd a_{n}...\dd a_{1} f(a)=\det A \int\prod_{i} \dd b_{n}...\dd b_{1} F(b),
\ee
at variance with the commuting case in which the factor on the right hand side would have been $\det^{-1} A$.
We define $\int\DD[a,\bar a]=\int\prod_{i} \dd a_{i}\dd \bar a_{i}$ the Grassmann measure.
In the multidimensional integral, the symbols $\dd a_{1}, ...,\dd a_{N}$ are again anticommuting with each other. The basic expression of the Grassmann analysis concern the Gaussian fermionic integrals \cite{samuel1980use1} which is related to the determinant
\bb
\det A=\int\DD[a,\bar a] \exp\Big(\sum_{i,j=1}^{N}a_{i}A_{ij}\bar a_{j}   \Big),
\ee
where $\{a_{i}, \bar a_{i}\}$ is a set of completely anticommuting Grassmann variables, the matrix in the exponential is arbitrary. The two Grassmann variables $a_{i}$ and $\bar a_{i}$ are independent and not conjugate to each other, they can been seen as composante of a complex Grassmann variables. The Gaussian integral of the second kind is related to the Pfaffian of the associated skew-symmetric matrix
\bb
\Pf A=\int \DD[a] \exp\Big(\frac{1}{2}\sum_{i,j=1}^{N}a_{i}A_{ij}a_{j}   \Big).
\ee
The pfaffian form is a combinatorial polynomial in $A_{ij}$, known in mathematics for a long time. The pfaffian and determinant of the associated skew-symmetric matrix are algebraically related by $\det A=(\Pf A)^{2}$. This relation can be most easily proved in terms of the fermionic integrals. The linear superpositions of Grassmann variables are still Grassmann variables and it is possible to make a linear change of variables in the integrals. The only difference with the rules of the common analysis, is that the Jacobian will now appear in the inverse power. New variables of integration can be introduced, in particular, by means of the transformation to the momentum space. The permanent of $A$ and the so-called haffnian can be written with Grassmann variables as well
\bb
&&{\rm perm} A=\int\DD[b,\bar b]\int\DD[a,\bar a] \exp\Big(\sum_{i,j=1}^{N}a_{i}\bar a_{i}A_{ij}b_{j}\bar b_{j}   \Big),\nn\\
&&{\rm hf} A=\int\DD[a,\bar a] \exp\Big(\frac{1}{2}\sum_{i,j=1}^{N}a_{i}\bar a_{i}A_{ij}a_{j}\bar a_{j}   \Big),
\ee
which are connected by the formula ${\rm perm} A=({\rm hf }A)^{2}$. We recall that the definition of the permanent differs from that of the determinant in that the signatures of the permutations are not taken into account.

\section{Plechko mirror symmetry}\label{appendixplechko}

\begin{figure}[tb]
\center{
\includegraphics[scale=1.2]{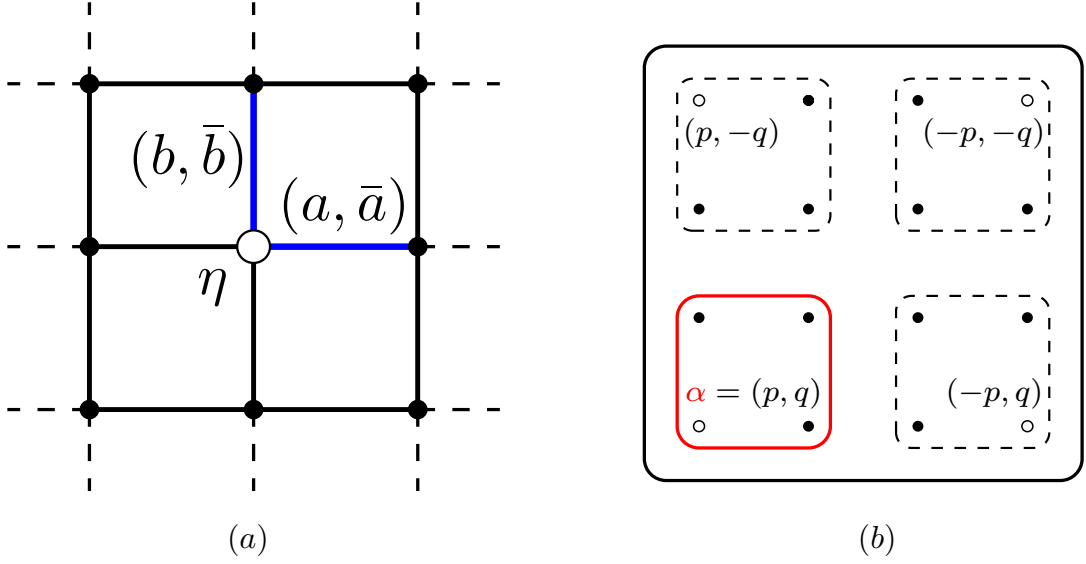}}
\caption{($a$) Grassmann representation for each dimer, with one nilpotent
variable $\eta$ per site, and two pairs of Grassmann variables for the two 
directions along the links connecting two neighboring sites.
($b$) Block partition of the Fourier 
modes. The modes 
considered in the summation \eref{S0} belongs to the sector inside the 
reduce domain (red delimitation) $1\le p,q \le L/2$. For one point labeled 
$\alpha=(p,q)$ inside this domain correspond 3 others points related by symmetry $p\rightarrow 
L+1-p$ and $q\rightarrow L+1-q$ (open circles).}
\label{block_even}
\end{figure}

In this appendix we briefly recall the method of resolution of the $2d$ dimer model based on the integration over Grassmann variables and factorization principles for the partition function introduced in the context of $2d$ Ising model \cite{plechko85a}. The general partition function for a graph with $N$ vertices
\bb
Q_0&=&\int\DD[\eta] \exp\Big(\frac{1}{2}\sum_{i,j=1}^{N}\eta_{i}A_{ij}\eta_{j}\Big),
\ee
can be written, for a square lattice of size 
$L\times L$ with $L$ even, as 
\bb\label{Q0def}
Q_{0}=\int\DD[\eta] \prod_{m,n}^{L}
(1+t_x\eta_{mn}\eta_{m+1n})(1+t_y\eta_{mn}\eta_{mn+1}),
\ee
where $\eta_{mn}$ are nilpotent and commuting variables on every vertices of the square lattice. The integrals can be done if we introduce a set of Grassmann variables $(a_{mn},\bara{mn},b_{mn},\barb{mn})$, ({\it cf.} \efig{block_even}($a$)), such that
\bb\nn
& &(1+t_x\eta_{mn}\eta_{m+1n})=\int\DD[\bar{a}]\DD[a]e^{a_{mn}\bara{mn}}(1+a_{mn}\eta_{mn})
(1+t_x\bara{mn}\eta_{m+1n}),
\\ 
& &(1+t_y\eta_{mn}\eta_{mn+1})=\int \DD[\bar{b}]\DD[b]e^{b_{mn}\barb{mn}}(1+b_{mn}\eta_{mn})
(1+t_y\barb{mn}\eta_{mn+1}).
\ee
This decomposition allows for an integration over variables $\eta_{mn}$, after 
rearranging the different link variables $A_{mn}:=1+a_{mn}\eta_{mn}$, $\bar A_{m+1n}:=1+t_x\bara{mn}\eta_{m+1n}$,
$B_{mn}:=1+b_{mn}\eta_{mn}$ and $\bar B_{mn+1}:=1+t_y\barb{mn}\eta_{mn+1}$. Then the partition function
becomes 
\bb
{Q}_0=\Tr{a,\bar a,b,\bar b,\eta}\prod_{m,n}^L(A_{mn}\bar 
A_{m+1n})(B_{mn}\bar B_{mn+1}),
\ee
where we use the notation for the measure of integration
\bb
\Tr{a,\bar a,b,\bar b,\eta}{X}(a,\bar a,b,\bar b,\eta)=
\int 
\DD[\bar{a}]\DD[a]\DD[\bar{b}]\DD[b]\DD[\eta] \prod_{mn}e^{a_{mn}\bara{mn}+b_{mn}\barb{mn}} X(a,\bar a,b,\bar b,\eta).
\ee
Then, the non-commuting link variables are moved in such a way that each 
$\eta_{mn}$ is isolated and can be integrated
directly. This rearrangement is possible in two dimensions thanks to the mirror 
ordering introduced by Plechko for the Ising model. The 
ordering process can be detailed as follow
\bb\nn
\prod_{m,n}^L(A_{mn}\bar A_{m+1n})(B_{mn}\bar B_{mn+1})
&=&\prodd{n=1}{L}(A_{1n}\bar A_{2n})(B_{1n}\bar B_{1n+1})(A_{2n}\bar A_{3n})(B_{2n}\bar B_{2n+1})\cdots
\\ \nn
&=&\prodd{n=1}{L}(A_{1n}\bar A_{2n})
(A_{2n}\bar A_{3n})\cdots(B_{1n}B_{2n}\cdots\bar B_{2n+1}\bar B_{1n+1})\nn\\
&=&\prodd{n=1}{L}
(B_{1n}(A_{1n}\bar A_{2n})B_{2n}(A_{2n}\bar A_{3n})\cdots\bar 
B_{2n+1}\bar B_{1n+1})\nn\\ 
&=&\prodd{n=1}{L}(\bar B_{Ln}\cdots \bar B_{2n}\bar B_{1n})(B_{1n}A_{1n}\bar 
A_{2n}B_{2n}A_{2n}\bar A_{3n}\cdots \bar A_{Ln}B_{Ln}A_{Ln}),
\ee 
where the products are ordered according to the orientation of the arrows. The 
Grassmann terms in brackets $(\cdots )$ on the first line of the previous 
equation are commuting objects, since they are integral representations 
of commuting scalars. This also imposes the boundary conditions $\bar 
A_{1n}=1$, $\bar A_{L+1n}=1$, $\bar B_{m1}=1$, and $\bar B_{mL+1}=1$, or 
$\bara{0n}=\bara{Ln}=\barb{m0}=\barb{mL}=0$ (for open boundary conditions 
only). We finally obtain the following exact expression 
\bb\label{Q0ord}
{Q}_0=\Tr{a,\bar a,b,\bar b,\eta}
\prodd{n=1}{L}\Big (\prodg{m=1}{L}\bar B_{mn}\prodd{m=1}{L}\bar 
A_{mn}B_{mn}A_{mn}\Big ).
\ee
The integration over the $\eta_{mn}$ variables is performed exactly, 
recursively from $m=1$ to $m=L$ for each $n$. Each integration leads to
a quantity $L_{mn}=a_{mn}+b_{mn}+t_x\bara{m-1n}+(-1)^{m+1}t_y\barb{mn-1}$ which 
is moved to the left of the products over $m$, hence a minus sign is needed in front of $\bar b$ each time a $L_{mn}$ crosses 
the product of $\bar B$ terms on the left. 
Finally
\beq
{Q}_0=\Tr{a,\bar a,b,\bar b}\prodd{m,n}{L}L_{mn},
\eeq
becomes an integration over products of linear Grassmann terms. This can be further
simplified by introducing additional Grassmann variables $c_{mn}$ such that 
\beq
L_{mn}=\int \dd c_{mn}\exp(c_{mn}L_{mn}).
\eeq 
This expresses ${Q}_{0}$ as a Gaussian integral over variables $(a,\bar a,b,\bar 
b,c)$, 
and therefore ${Q}_0$ is a simple determinant of a quadratic form. Indeed, after 
partially integrating over variables $(a,\bar a,b,\bar b)$ and symmetrization of the expressions, one obtains
\bb\label{Q0int}\nn
{Q}_0&=&\Tr{a,\bar a,b,\bar b,c}\exp\left 
(\sum_{mn}c_{mn}L_{mn}\right )\\ \nn
&=&\int\DD[c]\exp\sum_{mn}\left [\ff t_x(c_{m+1n}c_{mn}-c_{m-1n}c_{mn})+\ff t_y(-1)^{m+1}(c_{mn+1}c_{mn}-c_{mn-1}c_{mn})
\right ]\nn\\
&=&\int\DD[c]\exp {S}_0.
\ee
The computation of the determinant of this quadratic form can be done simply 
using Fourier transform satisfying open boundary conditions 
\bb\label{FourierC}
c_{mn}=\frac{2i^{m+n}}{L+1}\sum_{p,q=1}^Lc_{pq}\sin\left (\frac{\pi 
pm}{L+1}\right )\sin\left (\frac{\pi qn}{L+1}\right ),
\ee
with $c_{0n}=c_{L+1n}=c_{m0}=c_{mL+1}=0$. Inserting \eref{FourierC} into 
\eref{Q0int}, and using two following sum identities
\bb
\frac{2}{L+1}\sum_{m=1}^L\sin\left (\frac{\pi pm}{L+1}\right )
\sin\left(\frac{\pi qm}{L+1}\right )&=&\delta_{p,q},\\ 
\frac{2}{L+1}\sum_{m=1}^L(-1)^{m+1}\sin\left (\frac{\pi pm}{L+1}\right )
\sin\left(\frac{\pi qm}{L+1}\right )&=&\delta_{p+q,L+1},
\ee
we can finally put \eref{Q0int} into a block form of 4 independent Grassmann 
variables
\bb \label{S0}
{S}_0=\sum_{p,q}^{L/2}2it_x\cos\frac{\pi p}{L+1}\left 
(c_{pq}c_{-p-q}+c_{p-q}c_{-pq}\right )
+2it_y\cos\frac{\pi q}{L+1}\left (c_{pq}c_{p-q}+c_{-pq}c_{-p-q}\right ),
\ee
where, for example, $c_{-pq}$ is a short notation for $c_{L+1-pq}$. The 
summation is performed only for 1/4 of the Fourier modes, (see \efig{block_even}($b$)),
since the other are related by the symmetry $p\rightarrow L+1-p$ and 
$q\rightarrow L+1-q$. This block representation is convenient for computing the remaining integrals 
over the momenta, as a product of cosine functions as found by Kasteleyn Temperley and Fischer
\bb
{Q}_0=\prod_{p,q=1}^{L/2}\left [ 4t_x^2\cos^2\frac{\pi 
p}{L+1}+4t_y^2\cos^2\frac{\pi q}{L+1} \right ].
\ee
The main ingredient of this recipe is the mirror factorization \eref{Q0ord} of the partition function, which allows a direct integration over nilpotent variables. This mirror factorization is then the major obstacle to the generalization of this formalism in $d> 2$. Indeed, for the $3d$ case, a generalization of this factorization is far from obvious and remains to find.

\section{General monomer-dimer partition function}\label{dimermonomer}

\begin{figure}[h!]
\begin{center}
\includegraphics[scale=1]{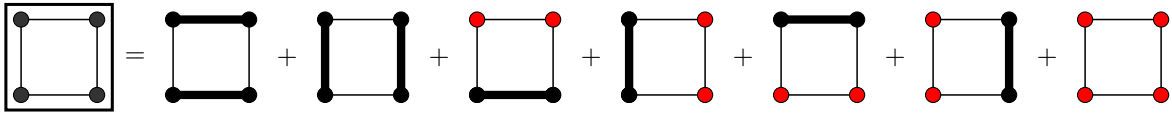}
\caption{Number of configurations of the general monomer-dimer model for a $2\times2$ square lattice.}
\label{default3}
\end{center}
\end{figure}

The general monomer-dimer problem is a much more complex and challenging problem in statistical physics and combinatorics, because the position of the monomers are not fixed either than their number ({\it cf.} \efig{default3} for $L=2$). From the point of view of theoretical physics, the number of monomers divided by the number of occupied site defines the monomer density $\rho$. It is long known that the full phase diagram of the monomer-dimer model does not admit any phase transition for $\rho>0$ \cite{heilmann1972theory,heilmann2004theory}. Furthermore the behavior of monomer-monomer correlations for finite density has been studied numerically \cite{krauth2003pocket}, and strong evidences for exponential correlations has been established, in accordance with mean-field calculations using Grassmann variables \cite{papanikolaou2007quantum}. From a computational point of view, the problem has been shown to belong to the $\#P$-complete enumeration class \cite{jerrum1987two} and all the methods available are either efficient but approximative \cite{baxter1968dimers,kenyon1996approximating} or exact but desperately slow \cite{ahrens1981paving}. In this short appendix one shows how to use our exact solution to express the partition function of this  enumerative problem.
Let us start by counting the number of ways $N_{2p}(M,N)$ of choosing the positions of $2p$ monomers on a $M\times N$ lattice, the result is a simple binomial expression $N_{2p}(M,N)=\binom{M^{2}/2}{p}\binom{N^{2}/2}{p}$. Using this formula we can sum up over the number of monomers $2p$ to obtain the number of ways to choose the positions of the monomers. Finally the number of terms in the full partition function is one (the pure dimer model) plus all the terms with an even number of monomers
(one choose $M=N=L$ for simplicity)
\beq
N(L)=1+\sum_{p=1}^{L^{2}/2}N_{2p}(L)=\frac{2^{L^{2}} \Gamma\big(\frac{L^{2}+1}{2}\big)} {\sqrt{\pi}\Gamma\big(\frac{L^{2}+2}{2}\big)}.
\eeq
This number grows as $2^{L^{2}}$ when the size of the lattice goes to infinity, making the problem impossible to solve analytically. Since our method allows to calculate exactly the partition function of the the dimer model on a square lattice of size $M\times N$ with an arbitrary even number of monomers, then we can formally write down the full monomer-dimer partition function as a sum over the number and the positions of monomers
\bb\label{full}
\mathcal{Z}=Q_{0}+\sum_{\{r_{i}\}}Q_{2}+\sum_{\{r_{i}\}}Q_{4}+...
\ee
which becomes simpler in the boundary case
\bb
\mathcal{Z}=Q_{0}\Big[1+\sum_{ij}C_{ij}+\sum_{ijkl}(C_{ij}C_{kl}\pm{\rm permutation})+\sum_{ijklmn}(C_{ij}C_{kl}C_{mn}\pm{\rm permutation})+...\Big].
\ee
The general formula \eref{full} allows for the numerical computation of the full partition function for small system sizes up to $L=8$.
Unfortunately, our method belong to the second category, the algorithm time grows exponentially with the size of the system.
\begin{table}[h!]\center
\begin{tabular}{|c|cccc|} \hline
$M\backslash N$ & ~$2$~ & ~$4$~   & ~$6$~  & ~$8$~  \\ \hline
$2$   & $7$ & $71$  & $733$ & $7573$                           \\
$4$   & $71$   & $10012$  & $1453535$ & $211351945$                            \\
$6$   & $733$   & $1453535$  & $2989126727$ & $61582117253688$                             \\ 
$8$   & $7573$   & $211351945$  & $6158217253688$ & $179788343101980135$       \\ \hline
\end{tabular}
\caption{Number of configurations of the general monomer-dimer model for a $L\times L$ square lattice computed using \eref{full}.}
\end{table}

\end{appendices}

\newpage

\bibliography{grassmann}

\end{document}